\documentclass[acmsmall,screen]{acmart}

\setcopyright{acmcopyright}
\copyrightyear{2023}
\acmYear{2023}
\acmDOI{}

\acmJournal{TOSEM}
\acmVolume{1}
\acmNumber{1}
\acmMonth{4}
\usepackage{caption}
\usepackage{subcaption}
\usepackage{multirow}
\usepackage{amsmath}
\usepackage{graphicx}

\usepackage{amssymb}
\usepackage{tabularx}
\usepackage{colortbl}
\usepackage{enumitem}
\usepackage[most]{tcolorbox}
\usepackage{csquotes}
\usepackage{fontawesome}
\usepackage{tikz}
\usepackage{tasks}
\newcommand{\todo}[1]{}
\renewcommand{\todo}[1]{{\color{red} TODO: {#1}}}

\definecolor{signif}{RGB}{228, 245, 246}
\definecolor{greyb}{RGB}{238, 238, 238}
\definecolor{greya}{RGB}{244, 244, 244}
\definecolor{signifl}{RGB}{229, 229, 200}
\definecolor{experience}{RGB}{236, 233, 237}
\definecolor{health}{RGB}{222, 232, 252}

\def\mybarhhigh#1#2{%%
   {\color{black}\rule{#1mm}{5pt}}  #2}

\usepackage{enumitem}

\newcommand{\boxquote}[1]{
\vspace{0.3em}
\hspace{-1em}
        \tcbox[on line, 
        boxsep=2pt, left=6pt,right=6pt,top=1pt,bottom=1pt,
        colframe=white,colback=gray!8]{%
            \parbox{0.93\linewidth}{%
            \vspace{0.3em}
            \hspace{0.2em}\faUsers
\textit{ " #1"}
                \vspace{0.2em}
            }
        }\vspace{0.2em}
}%
\definecolor{darkgreen}{rgb}{0.0, 0, 0.0} 
\definecolor{darkblue}{rgb}{0.0, 0.0, 0} 
\newcommand{\surveyquote}[1]{
\vspace{0.3em}
\hspace{-1em}
        \tcbox[on line, 
        boxsep=2pt, left=6pt,right=6pt,top=1pt,bottom=1pt,
        colframe=white,colback=gray!8]{%
            \parbox{0.93\linewidth}{%
            \vspace{0.3em}
            \hspace{0.2em}\faListAlt
\textit{ " #1"}
                \vspace{0.2em}
            }
        }\vspace{0.2em}
}%

\newcommand{\memo}[1]{
\vspace{0.5em}
\hspace{1em}
        \tcbox[on line, 
        boxsep=3pt, left=10pt,right=10pt,top=2pt,bottom=2
        pt,
        colframe=white,colback=gray!10]{%
            \parbox{0.90\linewidth}{%
            \vspace{0.5em}
            \hspace{0.5em}
            #1
                \vspace{0.5em}
            }
        }\vspace{0.5em}
}%

\newtcbtheorem{Summary}{\bfseries Summary}{enhanced,
  coltitle=black,
  top=0.15in,
  colframe=white,
  colback=gray!10!white,
  attach boxed title to top left=
  {xshift=0.5em,yshift=-\tcboxedtitleheight/2},
  boxed title style={size=small,colback=gray!40!white,colframe=gray!40!white}
}{summary}

\usepackage[]{mdframed}
\definecolor{steBoxLine}{rgb}{0.0, 0.0, 0.0}
\mdfdefinestyle{stebox2}{%
linecolor=steBoxLine,linewidth=2pt,%
leftmargin=0cm,rightmargin=0cm,%
roundcorner=5pt,
topline=false,bottomline=false,rightline=true,leftline=true,%
backgroundcolor=gray!05
}

\begin{document}

\title{A Proxy‑Stakeholder Approach to Requirements Engineering for Inclusive Navigation}

\author{Wei Wang}
\email{wei.wang5@monash.edu}
\affiliation{%
  \institution{Monash University}
  % \streetaddress{Wellington Rd}
  \city{Melbourne}
  % \state{Victoria}
  \country{Australia}
  % \postcode{3031}
}

\author{Anuradha Madugalla}
\affiliation{%
\institution{Deakin University}
  % \streetaddress{Wellington Rd}
  \city{Melbourne}
  % \state{Victoria}
  \country{Australia}
  % \postcode{3031}
}
  \email{anu.madugalla@monash.edu}

\author{John Grundy }
\affiliation{%
\institution{Monash University}
  % \streetaddress{Wellington Rd}
  \city{Melbourne}
  % \state{Victoria}
  \country{Australia}
  % \postcode{3031}
}
  \email{john.grundy@monash.edu}

\author{Paul McIntosh }
\affiliation{%
\institution{RMIT University}
  % \streetaddress{Wellington Rd}
  \city{Melbourne}
  % \state{Victoria}
  \country{Australia}
  % \postcode{3031}
}
  \email{paul.mcintosh@rmit.edu.au}

  \author{Charmine E. J. Hartel}
\affiliation{%
\institution{Monash University}
  % \streetaddress{Wellington Rd}
  \city{Melbourne}
  % \state{Victoria}
  \country{Australia}
  % \postcode{3031}
}
  \email{charmine.hartel@monash.edu}
  
\renewcommand{\shortauthors}{Wei et al.}

%%
%% The abstract is a short summary of the work to be presented in the
%% article.
\begin{abstract}
\color{darkblue}
Wayfinding, or the ability to navigate one's surroundings, is crucial for independent living and requires a complex combination of cognitive abilities, environmental awareness, and technology to manage this successfully. Individuals with cognitive impairment (IwCI) often face significant challenges in learning and navigating their environment. Despite its importance, mainstream navigation technologies are rarely designed with their diverse needs in mind. This study reframes the search for places as a socially distributed task and emphasizes the role of \textbf{\textit{proxy stakeholders}}, who act on behalf or in coordination with IwCI during navigation. Using a qualitatively led mixed-methods approach, which includes an international survey and a three-stage interview study, we examine the real-world strategies that proxy stakeholders employ to support daily navigation. The findings are synthesized into a set of empirically grounded design recommendations that emphasize customisability, collaborative use, and support for routine-based navigation. Our findings highlight key challenges and adaptive practices, which are synthesized into design recommendations that prioritize customisability, routine-based navigation, and multi-user coordination. By introducing the proxy stakeholder concept into the software engineering literature, we propose a more inclusive approach to requirements elicitation and offer practical guidance for designing navigation technologies that better reflect the complex realities of cognitive support.
\end{abstract}

%%
%% The code below is generated by the tool at http://dl.acm.org/ccs.cfm.
%% Please copy and paste the code instead of the example below.
%%

\begin{CCSXML}
<ccs2012>
   <concept>
       <concept_id>10003456.10010927.10003616</concept_id>
       <concept_desc>Social and professional topics~People with disabilities</concept_desc>
       <concept_significance>500</concept_significance>
       </concept>
   <concept>
       <concept_id>10002951.10003227.10003236.10003101</concept_id>
       <concept_desc>Information systems~Location based services</concept_desc>
       <concept_significance>500</concept_significance>
       </concept>
   <concept>
       <concept_id>10011007.10011074.10011075.10011076</concept_id>
       <concept_desc>Software and its engineering~Requirements analysis</concept_desc>
       <concept_significance>500</concept_significance>
       </concept>
   <concept>
       <concept_id>10011007.10011074.10011075.10011078</concept_id>
       <concept_desc>Software and its engineering~Software design tradeoffs</concept_desc>
       <concept_significance>500</concept_significance>
       </concept>
 </ccs2012>
\end{CCSXML}

\ccsdesc[500]{Software and its engineering~Requirements analysis}
\ccsdesc[500]{Software and its engineering~Software design tradeoffs}
\ccsdesc[500]{Social and professional topics~People with disabilities}
\ccsdesc[500]{Information systems~Location based services}

%% Keywords. The author(s) should pick words that accurately describe
%% the work being presented. Separate the keywords with commas.
\keywords{Navigation Tools Usability, Adaptive Map Design, Personalized Navigation Systems, Cognitive impairments, Inclusive Navigation Systems, Wayfinding Technologies, Assistive Technology, Spatial Navigation, Digital Wayfinding}

%% A "teaser" image appears between the author and affiliation
%% information and the body of the document, and typically spans the
%% page.

%%
%% This command processes the author and affiliation and title
%% information and builds the first part of the formatted document.
\maketitle

\section{Introduction}
\color{darkgreen} 
Software engineering researchers have increasingly recognized that requirements engineering (RE) practices often do not adequately account for the needs of underrepresented or vulnerable user groups \cite{heumader2018requirements, ollerton2012ipar, lowdermilk2013user,koenig2011ausgrenzung,spradlin2012you,churchill2018putting}. These gaps are especially pronounced for individuals with cognitive impairments (IwCI). In this study, we use 'cognitive impairments' to refer to a range of neurodevelopmental and acquired conditions that may affect memory, attention, spatial reasoning, and decision-making. This includes individuals with autism spectrum disorder, intellectual disabilities, ADHD, and acquired brain injuries—groups commonly supported by caregivers and health professionals in community or assisted living settings. Traditional RE methods—such as direct self-report or abstract task reflection—can be difficult for many IwCI to carry out \cite{nuseibeh2000requirements}. As a result, systems designed to support this population often rely on incomplete, decontextualized, or misaligned requirements \cite{heumader2018requirements}. To address this gap, inclusive RE increasingly calls for the participation of not only direct users but also the broader social networks surrounding them \cite{muller2021stakeholders,pacheco2012systematic}. Previous work recognizes the importance of indirect stakeholders \cite{muller2021stakeholders, muller2022so}, those affected by the system without directly operating it. However, these accounts overlook a critical subgroup: individuals who are not only affected by a system but also actively represent, mediate, or support its use on behalf of the primary user.  Despite their centrality in many support settings, proxy stakeholders have not been systematically conceptualized or formalized within software engineering literature.

Navigation technologies offer a compelling case for illustrating the value of a proxy-informed RE approach. Wayfinding is crucial for personal autonomy \cite{farr2012wayfinding,cannella2005choice}, but for IwCI, it can pose significant challenges  \cite{prestopnik2000relations, schnitzler2015user, jamshidi2020wayfinding, kato2003individual}. For such individuals, navigating unfamiliar environments may lead to disorientation, anxiety, or dependence on others \cite{Bosch2017FlyingSA,yu2020maps, courbois2013wayfinding, davis2014patterns, farran2012useful, farran2016route, purser2015development}. Critically, navigation for this group is often inherently collaborative, involving caregivers or peers who interpret cues, manage uncertainty, and provide corrective support \cite{dalton2019wayfinding,zacharias2001path}. Because navigation behaviour arises from this shared, distributed context \cite{Bae2024WayfindingIP, yassin2021others}, proxy stakeholders possess indispensable insights that direct elicitation from the user alone cannot reliably capture. Navigation technologies like Google Maps are mainly designed for typical users, overlooking the specific needs of IwCI, pedestrian accessibility, and cognitive load \cite{nizomutdinov2021development,tannert2018disabled}. This reflects a broader limitation of the prevailing \textit{one-size-fits-all} design paradigm, where requirements engineering processes fail to account for the variability in user abilities, preferences, and environmental constraints \cite{reichenbacher2023adaptivity，courbois2013wayfinding, davis2014patterns, farran2012useful, farran2016route, purser2015development}. These differences require customized design, such as multimodal cue delivery, simplified route representations, or error-tolerant interaction design \cite{fickas2008route, liu2009customizing,liu2006indoor, gomez2015design, chang2010autonomous,hervas2013assistive, fickas2010travelers}. Although existing research has advanced our understanding of IwCI wayfinding performance \cite{beeharee2006natural, fickas2008route, liu2009customizing,liu2006indoor, gomez2015design, chang2010autonomous,hervas2013assistive,fickas2010travelers}, much research is carried out under laboratory conditions without engaging real IwCI, which cannot fully reflect the practical challenges encountered by IwCI \cite{yang2018parent,bailey2010using, budrionis2022smartphone}. As a result, the requirements generated often fail to reflect the complexity of practical navigation scenarios, including environmental unpredictability and proxy stakeholder decision-making \cite{yang2018parent}. Social factors, such as being in the presence or interacting with others, can affect a users' strategies and expectation \cite{dalton2019wayfinding,zacharias2001path}. In real-world settings, navigation decisions may be shaped by interactions with caregivers, support workers, or peers—stakeholders who hold critical domain knowledge but are frequently omitted from formal RE processes. This oversight is partly due to the methodological complexity of modeling collaborative decision-making and emergent interactions \cite{Bae2024WayfindingIP, yassin2021others}. 

While prior RE research has examined the experiences of vulnerable populations such as IwCI \cite{heumader2018requirements, ollerton2012ipar, lowdermilk2013user,koenig2011ausgrenzung,spradlin2012you,churchill2018putting}, it often overlooks proxy stakeholders—such as carers or support workers who act on behalf of users with limited communicative or cognitive capacities. This exclusion is concerning because IwCI often struggle with metacognitive insight, such as planning and evaluating their own behavior, and may also have limited communication abilities \cite{igier2022roles, nader2014self,hollomotz2018successful,mcfarland2024adaptive}. Disability research has long established the value of these stakeholder inputs in understanding autonomy, support needs, and contextual barriers \cite{neumann2000use, orobor2025cross}. However, these voices remain marginal in mainstream RE practice. This study addresses this gap by centering proxy stakeholder perspectives in a mixed-method RE process tailored for inclusive navigation systems. In doing so, it responds to growing calls within software engineering to embed social and contextual understanding into system design for vulnerable groups \cite{saha2023benefits, harrington2020forgotten}, and formalizes proxy stakeholders as a foundational construct in inclusive requirements elicitation. This paper makes the following key contributions:

\begin{enumerate}
    \item Introduces the concept of \textit{proxy stakeholders} to software engineering and formally distinguishes it from indirect stakeholders, offering a new lens for inclusive requirements elicitation.
    \item Offers a set of practical insights—structured across three stages—for identifying, engaging, and interpreting input from proxy stakeholders;
    \item Provides design recommendations grounded in empirical data to enhance accessibility, usability, and stakeholder coordination in navigation systems;
    \item Demonstrates a mixed-methods RE approach that integrates proxy stakeholder insights to capture requirements in a socially complex, multi-stakeholder domain; and 
    \item Lays a foundation for future research and inclusive design in navigation software for IwCI, supporting broader independence and accessibility for diverse user groups.
\end{enumerate}

The remainder of this paper is structured as follows. Section 2 details the key concepts and provides an overview of the study's motivation, and reviews existing research in this domain, identifying the gaps addressed by this study. Section 3 outlines our study methodology and Section 4 presents the findings of our surveys and interviews. Section 5 provides targeted recommendations for improving navigation systems for IwCI and their carers. Section 6 highlights potential limitations, and Section 7 concludes and outlines some open research issues.

\color{black}
\section{Background}
\subsection{Map navigation skills and tools}
In our daily lives, we consistently depend on navigation techniques, requiring us to interpret spatial information to determine (whether we realize it or not) our current location (within a building, neighborhood, city, etc.). Wayfinding encompasses various behaviors that allow an individual to recognize their current location and efficiently travel to a non-visible destination within the environment \cite{golledge2003human,kitchin2002cognition}. Numerous elements influence our ability and effectiveness to find our way, including both internal factors (such as age, sex, strategy types, sense of direction, comprehension, spatial skills, etc.) and external factors (such as building density, presence of notable landmarks, etc.) \cite{prestopnik2000relations, schnitzler2015user, jamshidi2020wayfinding, kato2003individual}. Effective wayfinding requires that a person be aware of their current location within the environment, identify the next intended destination, and figure out the means to reach it from their current position using any available resources \cite{lawton2010gender}.

Digital map software, such as Google Maps, Yelp, and Waze, have revolutionised wayfinding and navigation, playing a crucial role in improving people’s quality of life. However, there are several challenges that restrict the accessibility of these systems and the advantages they offer \cite{froehlich2019grand}. First, these platforms focus mainly on data related to road networks and points of interest, with a noticeable absence of details related to pedestrian infrastructure and physical accessibility. Furthermore, due to their visual format and dependence on gestures and mouse interactions, digital maps may not be accessible to certain users \cite{yu2020maps}. In addition, these often convey information in ways that aren't straightforward and contain perplexing instructions necessitating strong spatial orientation skills for comprehension \cite{hervas2013assistive}. 

Despite the variety of maps available to us, most generally adhere to a one-size-fits-all approach \cite{reichenbacher2023adaptivity}. Universal strategies do not consider the varied traits of mobile map users, including spatial abilities, literacy proficiency, mental conditions, disabilities, and map reading experience. Furthermore, users differ in their expertise and familiarity with maps and some may experience color vision impairments. Technologies for IwCI have had limited adoption, possibly due to the absence of design and validation centered around the user \cite{ienca2017intelligent}. Research has often emphasised the technological capabilities and design aspects, rather than the practicality and acceptance of these technologies for this user group \cite{bachle2018assistive}.

\subsection{Navigation skills for individuals with cognitive impairments }

Cognitive impairments include conditions that are congenital, such as Down Syndrome and intellectual and developmental disabilities, those acquired from traumatic brain injury or illness, such as aphasia or amnesia, those that develop naturally with age, such as Alzheimer’s disease, and those due to complex causes such as schizophrenia \cite{chang2010autonomous}. Being able to move independently is an essential skill for IwCI, including those with mild cognitive impairment, who are expected to reach a degree of autonomy similar to that of non-disabled individuals \cite{cannella2005choice}. This ability can drive the performance of daily tasks, promote physical activity, foster social connections, and improve overall quality of life \cite{slevin1998independent}. IwCI appear to struggle particularly with learning and finding their way around their surroundings, facing challenges in recalling information, acquiring new knowledge, making routine decisions, and maneuvering through constructed environments \cite{Bosch2017FlyingSA,yu2020maps}. 

Research in laboratory settings has indicated that multiple aspects of wayfinding might be deficient in IwCI \cite{courbois2013wayfinding, davis2014patterns, farran2012useful, farran2016route, purser2015development}.
For example, \citet{davis2014patterns} evaluated the navigational skills of young adults with Down Syndrome and typically developing children. The experiment participants first observed a straightforward path within a virtual setting featuring indoor hallways and subsequently had to navigate the path independently, with those with Down Syndrome committing more mistakes and remembering fewer landmarks along the path.
\color{darkgreen}

\subsection{Requirements elicitation for inclusive software design}
Requirements elicitation—the activity of discovering and capturing stakeholders’ needs—is a fundamental phase in software engineering, particularly within requirements engineering (RE) processes for greenfield and evolutionary development projects \cite{coughlan2002effective, rizk2016requirements}. Traditional RE techniques include user interviews, prototyping, group elicitation, and contextual inquiry \cite{nuseibeh2000requirements}. However, these methods often fail to address the needs of diverse user groups, especially those with cognitive impairment. They need special focus to include diverse perspectives, especially since many of these traditional methods can only engage with a limited number of users, due to time and resource constraints. Therefore, many of the systems of today are not developed or designed for people with cognitive disabilities.

A growing body of literature highlights that the user base for modern software systems is increasingly heterogeneous, spanning differences in age, culture, socio-economic background, and physical or cognitive abilities \cite{heumader2018requirements}. Despite this, most RE practices do not adequately support the participation of individuals with cognitive disabilities. These individuals often encounter significant barriers in articulating needs or navigating structured elicitation processes, resulting in software that does not meet their lived realities. There is also a longstanding prejudice in the field—that individuals with intellectual or cognitive impairments may not be reliable informants during interviews or surveys \cite{koenig2011ausgrenzung}. The questioning and exploration of a provided problem can be challenging for engineers, especially in a context where the individual has little power or support to expand or reframe the scope of a project \cite{spradlin2012you}. However, ignoring these perspectives can result in systems that reinforce exclusion. In contrast, focusing on underrepresented users can produce innovations that benefit all \cite{churchill2018putting}. For example, closed captions were originally introduced to support those with hearing impairments, but are now widely adopted in public settings and mobile use contexts \cite{hong2010dynamic}.

The disconnect between developers and end-users is further amplified by systemic gaps in training. Developers are often expected to elicit, interpret, and encode user needs without formal expertise in UX, participatory design, or inclusive research methods \cite{ovad2015teaching,grundy2020human}. At the same time, the demographic makeup of the software development community is often homogenous—typically young, well-educated, and technologically fluent \cite{ovad2015teaching}. This homogeneity can create a disconnect, making it difficult for developers to empathise with and incorporate the diverse, human-centric needs of users from different backgrounds and ability levels during the software engineering (SE) process. As \citet{grundy2020human} note, a lack of empathy or understanding of diverse needs can lead to poor design decisions, especially under tight time and resource constraints. Recent studies have also highlighted how crowd-based RE approaches often fail to engage users from marginalised groups \cite{grundy2024developers}. Online sources such as user reviews, bug reports, and feature requests do contain rich feedback, including those pertaining to accessibility needs \cite{eler2019android, tizard2021voice}, yet these mechanisms risk excluding users with limited digital access, language barriers, or cognitive disabilities \cite{tizard2022voice}.

To better support inclusive requirements elicitation—particularly for IwCI—researchers such as \citet{heumader2018requirements} advocate hybrid approaches that combine inclusive participatory action research \cite{ollerton2012ipar} with user-centered design methods \cite{lowdermilk2013user}. Although such methods can produce deeply contextualized and ethical designs, they are resource-intensive and difficult to scale. The challenges of participative development with people with cognitive impairments are well recognized and discussed in recent literature \cite{heumader2018requirements}. Another important consideration is that people with cognitive impairments might not always have insight into their own behaviors or the ability to express those experiences in a way that is easily understood by researchers. It is well established in the literature that people with intellectual disabilities or related cognitive impairments often face challenges in metacognitive insight—that is, their ability to monitor, evaluate, and articulate their own functional behavior is diminished relative to peers who develop typically \cite{igier2022roles, nader2014self}. At the same time, many individuals in this population encounter difficulties in expressive or receptive communication, which may limit the depth or clarity of self‑report in interviews or questionnaires \cite{hollomotz2018successful,mcfarland2024adaptive}.

\subsection{A Socially situated and proxy-informed approach to requirements engineering}
The social model of disability frames disability not as an individual deficit, but as a consequence of the interaction between people living with impairments and environments that are laden with physical, social, communicative, and attitudinal barriers \cite{barnes2019understanding,shakespeare2006social}. This view informs our approach to requirements analysis, particularly for assistive technologies designed for IwCI. From this perspective, designing effective navigation systems must go beyond functionality to consider how socio-environmental factors limit or enable meaningful access.

A crucial implication of this worldview is the need to rethink how stakeholders are identified and involved in requirements engineering. While stakeholders are commonly defined in software engineering as those who can affect or are affected by a system’s objectives \cite{sharp1999stakeholder}, existing RE practices often overlook indirect or less visible actors. For example, stakeholder identification (SI) is often approached as a straightforward task involving only direct users and developers, with limited emphasis on extended care networks or social infrastructures surrounding the user \cite{muller2021stakeholders}. \citet{pacheco2012systematic}'s systematic review of SI methods found that there is currently no robust, standardized approach to identify the full range of stakeholders during RE elicitation. This gap is especially problematic in domains such as disability support, where end-users often depend on caregivers, family members, or allied health professionals to access, interpret, and act on software-mediated information. Such indirect stakeholders are not only affected by the operation of these systems—they are integral to ensuring that the systems are used appropriately, safely and in alignment with the evolving needs of the user \cite{muller2022so}. However, there has been limited conceptual development of stakeholder types that are not only affected by the system but who speak, act, and decide on behalf of the intended user—a role we argue is distinct and critical in contexts involving cognitive or communicative impairments.

This is particularly true for individuals with cognitive impairments, who may experience different levels of independence and support requirements \cite{pradhan2018accessibility}. While many IwCI can learn to use technology independently, they often require additional scaffolding, ongoing guidance, and a support network that extends beyond the app itself \cite{czaja2016designing}. Incorporating insights from stakeholders' communities becomes essential in this context \cite{saha2023benefits, greenbaum1993pd}. These approaches recognize that designers and users often do not share the same lived experiences or epistemic perspectives \cite{harrington2020forgotten}. Numerous studies have highlighted the importance of including caregivers, therapists, and family members in the health (e.g., \cite{orobor2025cross,neumann2000use}) or HCI domain (e.g., \cite{eisapour2018participatory,hornof2017designing,dai2025envisioning} ). Their involvement ensures that the systems are not only technically functional but also contextually and socially usable. In such cases, the lines between user and stakeholder blur, and requirements must be co-constructed with all those who share responsibility for the success of the system. To support this rethinking, we introduce the concept of a \textbf{Proxy Stakeholder}:\textit{An individual who provides input, guidance, or decision-making during the design and use of a system on behalf of a person with limited ability to articulate or represent their own preferences due to cognitive, communicative, or contextual constraints. }

Studies have found that proxy assessments by caregivers often correlate well with actual abilities in daily functioning for people with cognitive impairments \cite{neumann2000use}. In fact, recent mobility studies have explicitly gathered caregivers’ perspectives to evaluate navigation and independence among people with dementia and intellectual disability \cite{orobor2025cross}. These proxy stakeholders—whether family members or allied health professionals—routinely serve as interpreters, enablers, and co-users of assistive systems and are thus positioned as critical informants in the elicitation of real-world requirements. To address the complexities of engaging IwCI in research on technology use, our study adopted a proxy approach after extensive consultations with domain experts, including an interior architectural designer specializing in neurodiversity spaces and an academic researcher with expertise in assistive technology development. These consultations highlighted that proxy stakeholders offer unique advantages, including longitudinal insights into evolving needs, practical knowledge of real-world implementation challenges, and the ability to articulate observed patterns of technology use across diverse contexts. This proxy-stakeholder approach represents a significant departure from traditional RE methods and offers a promising pathway for developing more inclusive and contextually appropriate navigation systems for IwCI.

\begin{figure}[h]
        \centering
        \includegraphics[width=\textwidth]{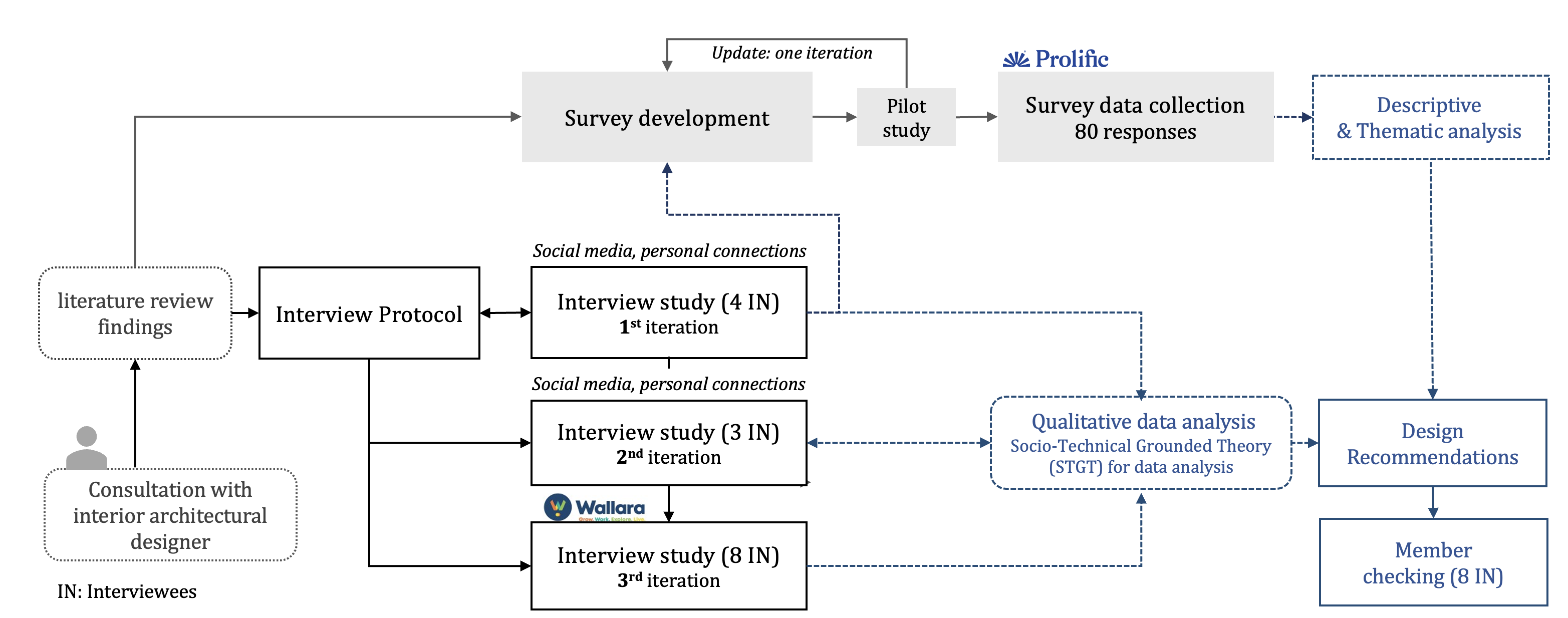}
        \caption{Study Methodology.}
        \label{fig:method}
    \end{figure}

\color{black}
\section{Methodology}
\color{darkblue}
We employ a \textbf{qualitatively led mixed
method} to enable us to explore various aspects of key challenges of navigation software for IwCI and their caregivers, and key requirements and design principles for more inclusive navigation software. Mixed methods research therefore heralds an opportunity for greater interdisciplinary collaboration \cite{hesse2015oxford}. Our design followed a \textbf{convergent parallel strategy} \cite{creswell2017designing}, in which qualitative and quantitative data were collected and analyzed concurrently but independently, then integrated during the interpretation phase. This strategy is especially suited for complex socio-technical phenomena where both breadth and depth are needed to understand user experiences and system design implications. As shown in Figure \ref{fig:method}, both the survey and interview components contributed iteratively and complementarily to the study. The study was started with a literature review that informed the development of the initial interview protocol and the development of the survey. Insights from the first wave of interviews further refined the content and structure of the survey instrument. In parallel, we conducted three iterations of in-depth interviews using evolving protocols informed by prior phases of data collection and analysis. 

Although survey and interview analyzes followed distinct methodological processes, they were integrated during the interpretation stage of the investigation. Recurring themes identified in the survey responses were triangulated with interview findings to strengthen the reliability of the identified challenges \cite{doyle2016overview}. In cases of divergence (e.g., differences in emphasis between support workers and family carers), we treated this not as a contradiction but as a signal of diverse stakeholder perspectives—an important consideration for inclusive design. This integrative synthesis informed the development of our design recommendations. Finally, we conducted a round of \textbf{member checking} with eight participants to validate the accuracy and relevance of the integrated findings. This mixed-methods strategy enabled us to combine scale (via survey) and depth (via interviews), supporting a nuanced, \textit{proxy stakeholder-informed} understanding of the design space for inclusive navigation systems. It also ensured that our findings were grounded in both empirical generality and contextual complexity, aligning with best practices in mixed-methods research \cite{doyle2016overview, fusch2018denzin, denzin2017research}.The study was approved by Monash University Human Research Ethics Committee (Project ID: 40245).

\subsection{Survey design}
\color{darkblue}
We wanted to gather a large amount of primary qualitative and some quantitative  data from IwCI and their carers. To do this, we chose to use an online survey. The design of this survey was informed by a comprehensive review of the existing literature, particularly those tailored to understand the navigation needs and experiences of IwCI \cite{gomez2015adapted,garcia2022survey,liu2009customizing,gomez2015design}. The goal was not to obtain a statistically representative sample, but to capture diverse and rich perspectives from those directly involved in supporting IwCI in their daily navigation and mobility. To achieve this, we employed a purpose-sampling strategy \cite{baltes2022sampling}, deliberately selecting participants who self-identified as caregivers or support professionals with direct experience assisting IwCIs. 

We used the Qualtrics survey platform and conducted the survey from October 2023 to January 2024, during which \textbf{80 responses} were collected in total. The survey consisted of closed and open questions, with some closed-ended questions using a Likert scale. Participants were recruited through the Prolific research platform \footnote{https://www.prolific.com/}, which allowed the targeting of individuals who met predefined inclusion criteria (i.e., current or recent caregiving experience for IwCI). The study used a rigorous two-stage process to ensure the validity of the data. In the first screening phase, prospective participants completed a detailed demographic questionnaire that evaluated: (1) duration of caregiving experience, (2) nature of their relationship with individuals with cognitive impairments (e.g., caregiver, healthcare professional), and (3) specific types of supported impairments. We have recorded 169 participants who show interest in our study. Only 115 respondents who demonstrated consistent, required experience through comprehensive responses were invited to participate in the main survey study. Among them, 80 completed the survey. The main survey includes a total of 22 questions which are divided into three sections: 1. Demographic questions, 2. Difficulties in navigation for IwCI, 3. Difficulties encountered when traveling with IwCI (Appendix \ref{app:stageone}). Although our sampling strategy ensured that we reached a relevant and experienced participant group, we recognize that the sample may not be fully representative of all caregivers for IwCI globally. This limitation and its implications for generalizability are discussed in the Threats to Validity section.

A pilot study was instrumental in refining the survey's design and focus, relying on the expertise of two key participants whose backgrounds provided insight into the realm of accessibility in software design and IwCI. The first participant is an academic with a focus on designing accessible software. The second participant is a nurse with some experience working with IwCI. Both participants provided feedback on the survey questions, which we used to revise the language and include additional choice options to improve clarity and comprehension. Following these improvements, we conducted our survey study.

\subsection{Interview study design}

To extend our literature findings and survey findings with deeper qualitative data, we chose to use semi-structured interviews. Our interview protocol was systematically designed through a comprehensive literature review of accessible navigation studies, incorporating validated question structures from previous work in cognitive accessibility research. Subsequently, the interview questions underwent a verification, review, and refinement process conducted by the second author, an expert with extensive experience in accessible navigation.
The interview process was conducted over three iterative waves to ensure methodological rigor and consistency (see Figure \ref{fig:method}). During the first iteration, interviews were jointly conducted by two co-authors to establish a shared understanding of the protocol and enhance data quality. The second author served as the primary facilitator, leading the discussion, while the first author acted as the moderator, documenting real-time observations and ensuring adherence to the interview guide. This dual-interviewer approach allowed immediate post-session debriefs and collaborative reflection, strengthening the reliability of early data collection. For subsequent iterations (second and third iterations), the first author independently conducted all interviews, maintaining consistency through structured protocols. 

Our interview study consisted of two phases of data collection. The first phase gathered demographic information and assessed the extent of participation in IwCI through a Qualtrics survey. The second phase involved in-depth interviews to capture participants’ perspectives on the difficulties IwCI face during outings, the responsibilities of caregivers, and the challenges encountered in providing support. Since our goal was to obtain context-rich experience-based insights rather than to obtain a statistically representative sample, we employed a hybrid purposive–convenience sampling strategy \cite{baltes2022sampling, de2015characterizing}. Specifically, purposive sampling was used to ensure that all participants had a direct and relevant caregiving experience, allowing the study to capture nuanced knowledge grounded in lived practice. This was complemented by convenience sampling, which leveraged accessible recruitment channels to efficiently reach eligible participants within the target population.

The interview process was conducted in \textbf{three distinct iterations} to ensure a diverse and comprehensive understanding of the phenomena under investigation. While we did not conduct a formal pilot study beyond the first iteration, this initial wave functioned as a formative stage in which the interview protocol was tested and refined in practice. In the first iteration, participants (n = 4) were recruited through personal connections and targeted social media advertisements, with the aim of rapidly engaging carers to pilot interview questions and identify emerging themes. Two co-authors conducted interviews together to align on protocol, document observations, and debrief, allowing for minor adjustments before the next rounds. In the next two iterations, individual researchers conducted interviews using the refined protocol. The second iteration (n = 3) extended recruitment to additional contacts within the research team’s networks, deliberately seeking individuals with varied caregiving roles, including family caregivers and healthcare professionals. In the third iteration (n = 8), we partnered with Wallara\footnote{https://www.wallara.com.au/}, a disability support organization, to reach a broader spectrum of carers. This partnership enabled us to access paid support workers with diverse experience supporting IwCI across multiple contexts, such as community outings and workplace environments. In total, our interviews involved a total of \textbf{15 interview participants}. Data saturation was monitored throughout the process. After the second iteration, recurring patterns emerged in participants’ accounts of navigation challenges, stakeholder roles, and preferred design features. The later stage of the third iteration confirmed these themes, producing minimal novel codes, indicating that thematic saturation had been reached. This supported our decision to conclude the recruitment of 15 participants.

\color{darkblue}
\subsection{Data analysis}

\subsubsection{Survey data analysis}

Our approach to analyzing survey data combined both \textbf{descriptive} and \textbf{thematic analysis} to generate a well-rounded understanding of the responses of the participants. Key descriptive statistics, including means, modes, and frequency distributions, illuminated general trends and patterns within the data, offering a preliminary understanding of the overarching tendencies in the navigation pattern of IwCI. To complement this, we applied \textbf{\textit{thematic analysis}} to explore the qualitative, open-ended responses in greater depth. Qualitative survey responses are brief and fragmented compared to interviews or focus groups. They reveal key themes, with questions targeting specific aspects like navigation challenges, communication needs, and emotional responses, resulting in responses clustered by question rather than dataset-wide themes. Thematic analysis was especially suitable in this context because it supports flexible, question-by-question coding while remaining sensitive to the context and diversity of individual responses \cite{braun2012thematic}. We adopted Braun and Clarke’s six-phase framework for thematic analysis \cite{braun2006using}, and the process unfolded as follows:
\color{darkgreen}

\begin{enumerate}[left=0pt,itemsep=2pt, topsep=5pt]
    \item Phase 1: Familiarization. we immersed in the data by conducting multiple close readings of all 80 open-ended responses. This process yielded initial analytical notes and observations about prevalent concepts for the open-end questions related the navigation related questions in the survey.
    \item Phase 2: Initial Coding. Our work in phase 2 led to the development of our understanding of the navigation process with IwCI and their potential preferences for existing features in navigation tools. This is being done by the first author. 
    \item Phase 3: Searching for Themes. The codes were clustered into potential themes through affinity diagramming involving two authors. We identified candidate themes which we mapped against the original data to verify their prevalence and coherence. 
    \item Phase 4: Reviewing Themes. We subjected the preliminary themes to testing with another author who has extensive experience in designing accessible navigation for various groups of users.
    \item Phase 5: Defining and Naming Themes. We considered ways to organize the properties and identify their relevance to the design of navigation applications for IwCI and their carers.
    \item Phase 6: Producing the Report.
    
\end{enumerate}
To ensure reliability, independent coding was initially performed by one researcher, followed by validation and iterative refinement by a second researcher.

\color{darkblue}
\subsubsection{Interview data analysis: Socio-Technical Grounded Theory for data analysis}

We utilized the data analysis techniques of \textbf{Socio-Technical Grounded Theory (STGT)} \cite{hoda2021socio} to examine the data collected from interviews. This choice was mainly influenced by two factors. First, the focus of our study closely corresponds to the principles of the socio-technical research framework that underpins STGT. Secondly, STGT enables selective application by incorporating its fundamental data analysis methods of open coding and constant comparison, along with memoing, within a mixed-method research approach. Traditional grounded theory methodologies \cite{glaser1992basics, strauss1994grounded} were designed as independent approaches for theory development. By employing STGT for data analysis, we embraced a comprehensive methodology, acknowledging the intricate and interconnected nature of the socio-technical phenomena being studied. Before transcribing the audio recordings, we obtained consent from the participants and subsequently managed and analysed the data using NVivo. 

\begin{figure}[h!]
        \centering
        \includegraphics[width=\textwidth]{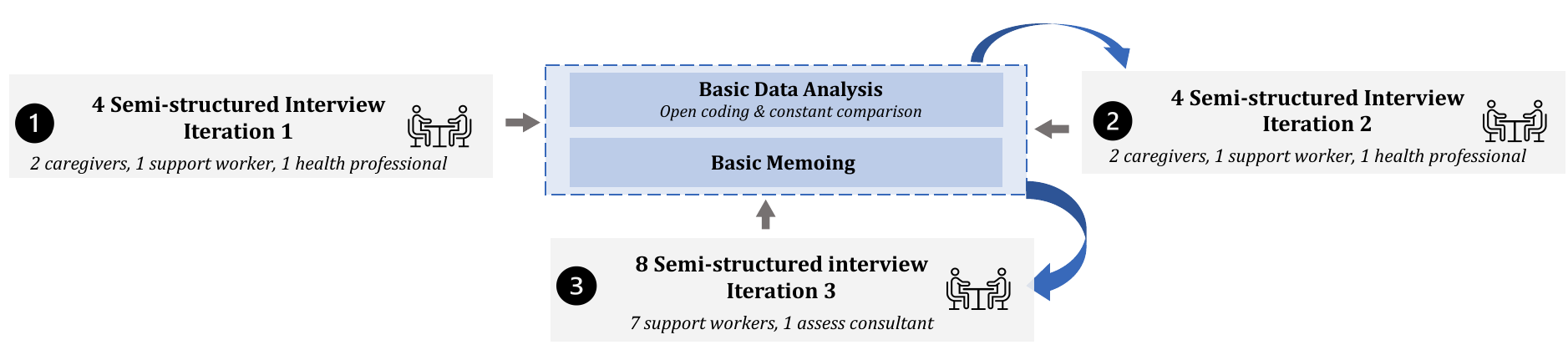}
        \caption{Process of applying STGT for data analysis in the interview study.}
        \label{fig:STGT}
    \end{figure}

The primary author conducted all interviews and initially coded the qualitative data for continuity and context. To improve reliability and minimize bias, a second researcher from the first interview wave reviewed the coding structure and engaged in discussions during the analysis. In Section \ref{interview}, we outline the concepts and categories derived from the STGT analysis. An example of the application of STGT for data analysis \cite{hoda2021socio} is detailed below:

\begin{enumerate}[left=0pt,itemsep=2pt, topsep=5pt]
    \item \textbf{Open Coding and Constant Comparison:} We analysed the audio transcripts and extracted various codes from the raw data. We provide some examples below.\\
    \textbf{Raw Quote:} \textit{ "We will verify today's temperature upon our arrival and gather information about the crowd or noise condition at the location. ”} \\
    \textbf{Code 1:} \textit{Providing sensory information on maps} \\

    \textbf{Raw Quote}: \textit{“Google Maps displays how busy various locations are, allowing us to plan our visit accordingly. It would be beneficial if they provided information on different sensory aspects as well."}\\
    \textbf{Code 2:} \textit{Specify the sensory intensity for various sensory stimuli}\\
    
    \noindent \textbf{Concept:} The two code examples given above suggest a concept: "Sensory information"
    
    \noindent \textbf{Category:} By analyzing the memos created during the coding process and recognizing the codes and concepts identified, we established several categories. The provided memo example demonstrates the category related to \textit{"accessibility and sensory information"}.

    \item \textbf{Sampling Strategy and Saturation} The data collection and analysis process followed an \textit{iterative and interleaved} method, as illustrated in Figure \ref{fig:STGT}. Our study started with \textbf{convenience sampling} , which involves selecting relevant participants that are easily available to the research team, due to limited access to health professionals, caregivers, and support workers who have experience with IwCI. We started our initial convenience sampling with the primarily caregivers and health professionals in the interview studies \textbf{(\textit{Iterations 1 and 2)}}. Once we had conceptualized the ideas from the interview data, we used \textbf{purposive sampling} to collect additional data, which helped us to clarify the concepts that were developing. For example, the majority of the interviewees in Iterations 1 and 2 were caregivers and health professionals. Consequently, we identified the need for refinement, leading us to enlist more participants from support workers in \textbf{	\textit{Iteration 3}} (see Figure \ref{fig:STGT}). \textit{Iteration 3} marked the point of saturation, saturation was systematically assessed through two concurrent measures: (a) no new substantive concepts emerging in final interviews, (b) full development of category properties. The final sample size (N=15) aligns with empirical studies of saturation in grounded theory research, where 12-20 participants typically suffice for well-bounded phenomena \cite{2018Data}. This was further confirmed through member checking, where participants validated the comprehensiveness of identified themes. Although larger samples might reveal additional nuances, our rigorous approach ensured conceptual depth while respecting practical constraints in recruiting specialized populations.

    \item \textbf{Memoing} played a crucial role in our approach, allowing us to explore emerging concepts and potential relationships between them, as outlined by \citet{hoda2021socio}. These memos were essential instruments for recording 	\textit{important insights and thoughts} derived from our open-coding activities.

    \end{enumerate}
    \memo{\textbf{Memo on "Accessibility and Sensory Information". } Participants have emphasised the need for additional details on the map to help IwCI utilise it more effectively. They proposed including information tailored to specific users, such as disability and sensory information and navigation progress. Besides providing user-relevant data, more insights about the navigation area would be beneficial, including details on the accessibility features like lifts, quiet rooms, exits, and entries. Additionally, there should be more information to confirm these details about the place. All this information is pertinent to accessibility and sensory information.}

    \subsubsection{Integration of the data analysis}
For a thorough grasp of the data gathered, \textbf{the analysis of the interview and survey data was integrated} into the reporting of this study. Following \citet{fetters2013achieving}'s joint display framework, we first conducted parallel analyzes where preliminary survey themes informed subsequent interview probes, while emergent interview insights prompted re-examination of survey responses for corroborating or contradictory patterns. Discrepancies between datasets were resolved by checking with participants of members, applying weighting criteria based on the prevalence of the data and the contextual richness. 

Some potential reason for divergence is acknowledged in our study since the methodological differences between the two phases of research; for example, the use of anonymous methods in a quantitative phase and non-anonymous methods in a qualitative phase might lead to different responses \cite{safer1997self}. Another explanation is that the survey may not be sensitive enough to pick up on complex experiences that have been reported qualitatively \cite{lee2015mixed}. Also, the participant group of the survey and interview study is a different cohort. 

 The nature of interviews inherently allows for a deeper exploration of individual experiences, capturing context, emotions, and unique challenges that may not be reflected in survey data alone \cite{WILSON201423}. The results are systematically categorized into \textit{navigation patterns} and \textit{navigation software design}, each further divided into subcategories to provide clarity and insights into specific areas. The final synthesis was achieved through visual joint displays that mapped quantitative survey frequencies to qualitative interview exemplars, creating hierarchical theme structures that connected macro-level survey patterns with micro-level interview experiences. Throughout our reporting, we maintain transparent provenance tracking: interviews-derived insights are marked with (\faUsers), survey-based evidence with ( \faListAlt) and convergent findings with both (\faUsers \faListAlt). 
 
This approach, grounded in \citet{denzin2017research} dialectical principles, preserves the complementary strengths of both methods while providing a robust empirical basis for our conclusions.
\begin{figure}[b]
    \centering
    \includegraphics[width=0.8\linewidth]{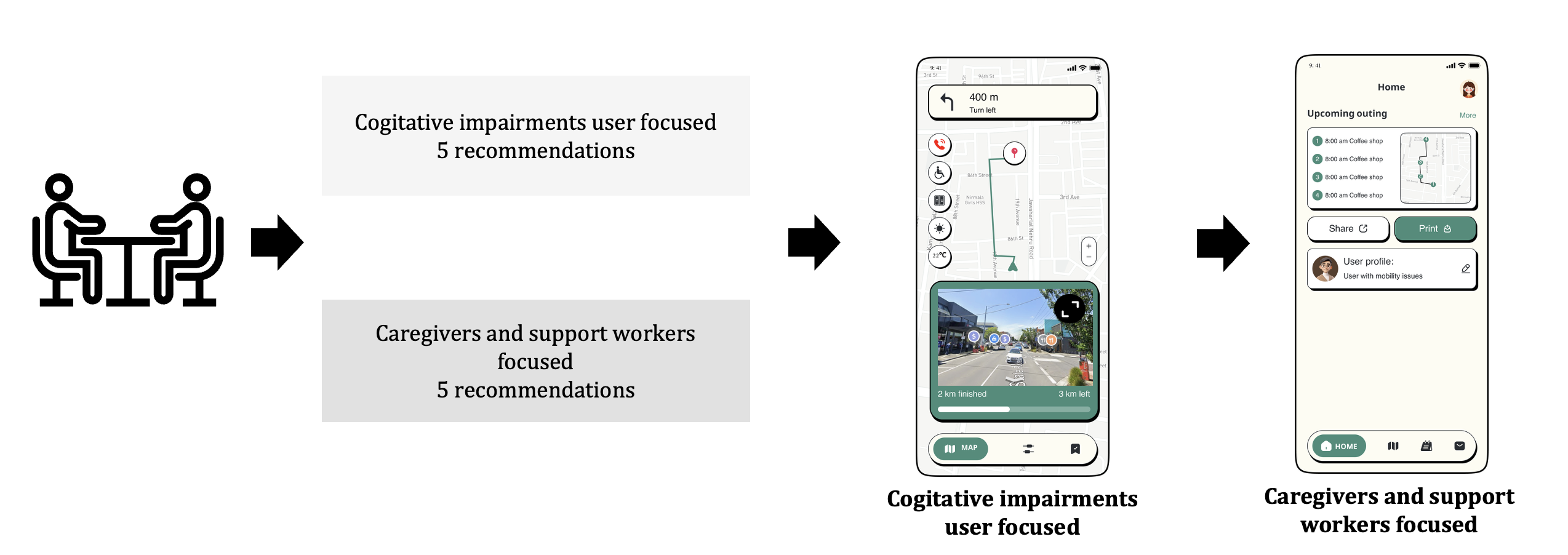}
    \caption{Member checking interview process}
    \label{fig:member}
\end{figure}

\color{darkgreen}
\subsection{Member checking}
Member checking, or participant validation, is a technique used in qualitative research to ensure the credibility of the results \cite{Birt}. This process addresses potential biases from researchers' personal beliefs of researchers by involving participants in the validation of the results \cite{Qualitativeresearching_2025, Qualitative}. We conducted follow-up interviews with eight participants who previously participated in our study, presenting the results obtained, particularly the design recommendations. During follow-up interviews, we presented a summary of the key findings, including the main themes and preliminary design recommendations. We also visualized several of these recommendations within a navigation prototype, allowing participants to see how the proposed designs could be applied in real-world contexts (see Figure \ref{fig:member}). Participants were invited to comment on the precision, clarity and relevance of these interpretations in relation to their lived experiences. The feedback from the sessions confirmed that the identified themes accurately reflected participants’ perspectives. Minor refinements were made to improve wording and emphasis in several recommendations, enhancing their clarity and practical alignment with participants’ contexts.

\color{black}
\section{Results}
\subsection{\faUsers \hspace{0.05cm} \faListAlt\hspace{0.05cm} Demographics of survey and interview participants }
\begin{table}[b]
\centering
\caption{Survey participants demographics information (n=80)}
\label{tab:surveydemographics}
{\scriptsize
\begin{tabular}{p{65mm}p{65mm}}
\hline
\toprule
    \begin{tabular}[t]{p{33mm}p{4mm}l}
    \textbf{Demographics} & \textbf{\#}  & \textbf{\% of Participants}  \\ \hline
    \multicolumn{3}{l}{\textit{\textbf{Gender}}}\\ \hline
         Female&  49& \mybarhhigh{9.8}{61\%}\\
        Male & 29 &\mybarhhigh{5.8}{36\%} \\
       Prefer not to tell &2& \mybarhhigh{0.4}{3\%}\\
        \multicolumn{3}{l}{\textit{\textbf{Age}}}\\ \hline
        18 - 24&17& \mybarhhigh{3.4}{21\%}\\
        25 - 34&34& \mybarhhigh{6.8}{43\%}\\
        35 - 44&19& \mybarhhigh{3.8}{24\%}\\
        45 - 54&7& \mybarhhigh{1.4}{9\%}\\
        55 - 64&3& \mybarhhigh{0.6}{4\%}\\

    \multicolumn{3}{l}{\textit{\textbf{Years of experience with cognitive impairment users}}}\\ \hline
    1-10&62& \mybarhhigh{12.4}{78\%}\\
    \multicolumn{3}{l}{\textit{Average:4, Median:3, Max:10, Min:1}}\\
    11-20&12& \mybarhhigh{2.4}{15\%}\\
    \multicolumn{3}{l}{\textit{Average:15, Median:15, Max:20, Min:1}1}\\
    21-30&3& \mybarhhigh{0.6}{4\%}\\
    \multicolumn{3}{l}{\textit{Average:24, Median:24, Max:25, Min:22}}\\
    More than 30&3& \mybarhhigh{0.6}{4\%}\\
    \multicolumn{3}{l}{\textit{Average:39, Median:42, Max:45 Min:31}}\\
    \end{tabular}
&
\begin{tabular}[t]{p{33mm}p{4mm}l}
    \textbf{Demographics} & \textbf{\#}  & \textbf{\% of Participants}  \\ \hline
    \multicolumn{3}{l}{\textit{\textbf{Country}}}\\ \hline
        UK and Northern Ireland&31& \mybarhhigh{6.2}{38\%}\\
        Portugal&10& \mybarhhigh{2}{12\%}\\
        South Africa&5& \mybarhhigh{1}{6\%}\\
        Italy&4& \mybarhhigh{0.8}{5\%}\\
        Australia&3& \mybarhhigh{0.6}{4\%}\\
        Greece&3& \mybarhhigh{0.6}{4\%}\\
        Spain&3& \mybarhhigh{0.6}{4\%}\\
        United States of America&3& \mybarhhigh{0.6}{4\%}\\
        \multicolumn{3}{p{60mm}}{Austria, Canada, Israel, Mexico, Poland,\textbf{ 2\%} Each}\\
        \multicolumn{3}{p{60mm}}{Belgium, Finland, Germany, Hungary, India, Latvia, Sweden, Switzerland, \textbf{1\% }Each}\\
    \multicolumn{3}{l}{\textit{\textbf{Roles in caring the cognitive impairment users }}}\\ \hline
    Health professionals&49& \mybarhhigh{9.8}{63\%}\\
    Support workers&15& \mybarhhigh{3}{19\%}\\
    Caregivers&10& \mybarhhigh{2}{13\%}\\
    Other&6& \mybarhhigh{1.2}{8\%}\\

    \end{tabular} \\
\bottomrule
\end{tabular}}
\end{table}
\color{darkgreen}
To aid in the interpretation of the findings, it is important to clarify the distinctions between caregivers and support workers within our sample of participants. While both groups provide support to IwCI, their roles differ significantly in terms of context, scope, and lived experience. Caregivers in this study primarily refer to unpaid family members who provide long-term, emotionally invested care. Their perspectives are shaped by intimate, day-to-day involvement with IwCI across personal and familial contexts. In contrast, support workers are typically paid professionals employed by disability service providers, working with multiple clients in structured environments such as group homes, community programs, or supported employment services. These differences often influenced how participants described the challenges of navigation: caregivers tended to focus on emotional stress, routine disruptions, and long-term wellbeing, while support workers emphasized procedural hurdles, safety concerns, and environmental design barriers. Recognizing these role-based nuances enriches our understanding of the diverse support ecosystems surrounding IwCI.

\color{black}
Our survey received 80 valid respondents and participants had a diverse demographic profile, summarized in Table \ref{tab:surveydemographics}. Most are women (61\%), and we have a significant spread across various age groups, notably within the 25-34 age bracket (43\%). Most of the participants (78\%) indicated having between 1 and 10 years of experience with IwCI, averaging 4 years within this range. Geographically, the United Kingdom emerged as the most common country of residence (38\%), followed by a variety of other countries, including Portugal, South Africa, and Italy, among others, showing an international spread. Professionally, health professionals constituted the largest group (63\%), supplemented by support workers (19\%) and caregivers (13\%), highlighting a wide spectrum of roles involved in the care of IwCI. Among health professionals, the distribution includes medical professionals and nurses (each representing 13\% and 19\%, respectively), followed by occupational therapists, physiotherapists, and psychologists (1\%, 6\%, and 6\%, respectively). 

\begin{figure}
    \centering
    \includegraphics[width=1\linewidth]{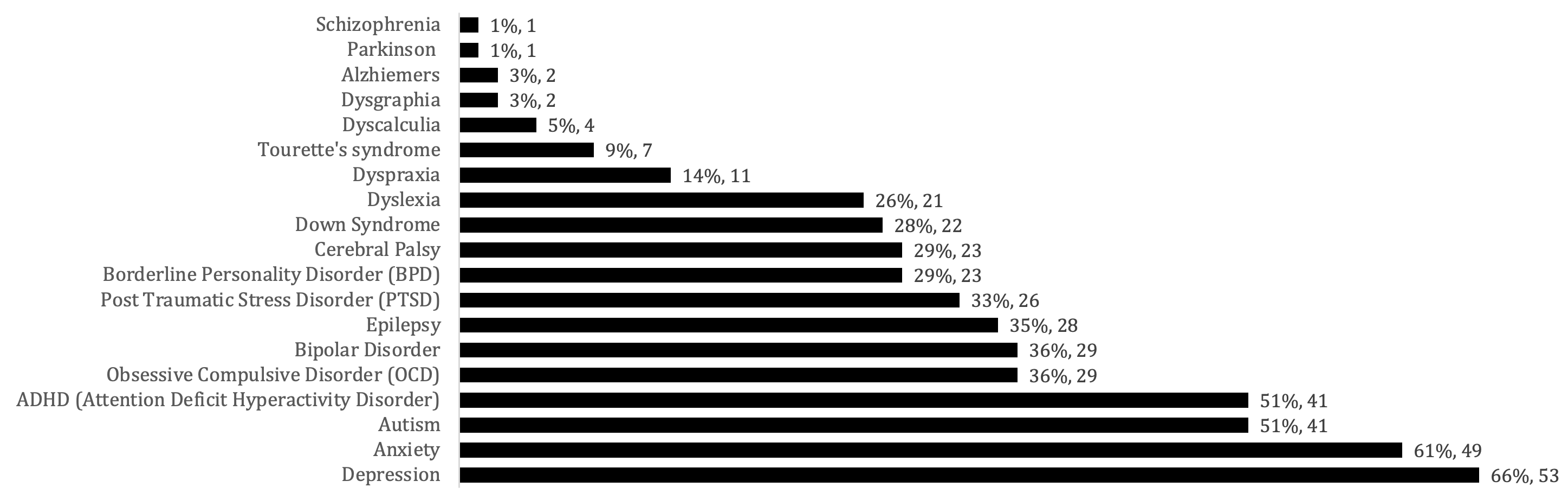}
    \caption{Different cognitive impairments of the individuals survey participants have cared for}
    \label{fig:surveycon}
\end{figure}

We also analyze the prevalence of various cognitive impairments that these carers are responsible for (see Figure \ref{fig:surveycon}). Depression and anxiety are very prevalent conditions that affect 66\% and 61\% of the participants, respectively. This highlights a notable overlap between cognitive impairments and mental health challenges, a well-documented connection in the literature \cite{castaneda2008review, roca2015cognitive}. Autism and ADHD also show a notable presence, each reported by more than half of the participants (51\%). Rarer conditions within the group include Dyspraxia (14\%), Tourette's syndrome (9\%), and Dyscalculia (5\%), with Dysgraphia, Alzheimer's, Parkinson’s, and Schizophrenia each reported by less than 5\% of participants. This distribution shows the vast spectrum of cognitive impairments, from the most common mental health issues (e.g, OCD, Depression and Anxiety) to less prevalent neurological (e.g., Parkinson, Alzheimer, Cerebral Palsy and Epilepsy) and developmental disorders (e.g., Dyspraxia, Dyslexia, Down Syndrome, Autism and ADHD), reflecting the complexity and variability of cognitive impairments in the population studied. 

\begin{table}[ht]
\centering
\caption{Participants demographic for interviews.}
\label{tab:codes_challenges}
  \resizebox{0.7\textwidth}{!}{%

\begin{tabular}{p{10mm}p{14mm}p{14mm}p{38mm}p{33mm}}

\toprule
 \textbf{\#} & \textbf{Age} & \textbf{Gender}  & \textbf{Role} &\textbf{Years of experience} \\

\hline
\multicolumn{5}{l}{\textit{\textbf{Iteration 1:} Each interview study took 50-60 minutes.}}
\\
\hline
IN 1&25 - 34&Female&Support workers&10 years\\
IN 2&65 - 74&Female&Caregivers&4 months\\
IN 3&35 - 44&Female&Health Professionals&15 years\\
IN 4&45 - 54&Female&Caregivers&25 years\\

\multicolumn{5}{l}{\textit{\textbf{Iteration 2:} Each interview study took 50-60 minutes.}}
\\
\hline
IN 5&35 - 44&Female&Health Professionals&15 years\\
IN 6&45 - 54&Female&Caregivers&15 years\\
IN 7&45 - 54&Female&Support workers&20 years\\

\multicolumn{5}{l}{\textit{\textbf{Iteration 3:} Each interview study took 40-50 minutes.}}\\
\hline
IN 8&35 - 44&Male&Support workers&8 years\\
IN 9&45 - 54&Female&Support workers&9 years\\
IN 10&25 - 34&Male&Support workers&1 years\\
IN 11&45 - 54&Female&Support workers&20 years\\
IN 12&25 - 34&Female&Access Consultant&10 years\\
IN 13&25 - 34&Male&Support workers&10 years\\
IN 14&25 - 34&Female&Support workers&1 year\\
IN 15&35 - 44&Female&Support workers&>5  years\\

\bottomrule
\end{tabular}}

% \resizebox{0.75\textwidth}{!}{%
% \tiny
% \begin{tabular}{r}
% The total per challenge is not the sum of the respondents since the participants often provided an answer that was categorized into \textbf{more than one challenge}.
% \end{tabular}
% }
% \vspace{-0.2cm}
\label{tab:demogr}
\end{table}

The demographic data from our interview study are summarized in Table \ref{tab:codes_challenges}. We have a cohort of participants with a concentration in the 25-34 and 45-54 age ranges, each constituting 33\% of the group, followed by the 35-44 age range at 27\%, with a single participant in the 65-74 age bracket. Most participants identify as female (80\%). Geographically, the majority are from Victoria, Australia, accounting for a significant 93\%, with only one individual from New South Wales. The roles of the participants are largely in support work (60\%), alongside caregivers and health professionals, each of which constitutes 13\% of the sample, while the remaining include roles such as access consultant. Experience levels among participants vary, with several indicating more than a decade of experience, and a few are relatively new, with as little as 4 months to a year. The cohort has experience caring for individuals with a variety of cognitive impairments, most commonly autism (87\%) and epilepsy (80\%), followed by ADHD, anxiety, depression and OCD among others. This demographic snapshot provides a foundation for understanding the diversity and depth of experience of those involved in the study.

\subsection{Navigation patterns of IwCI and carers}

\begin{figure}
    \centering
    \includegraphics[width=0.8\linewidth]{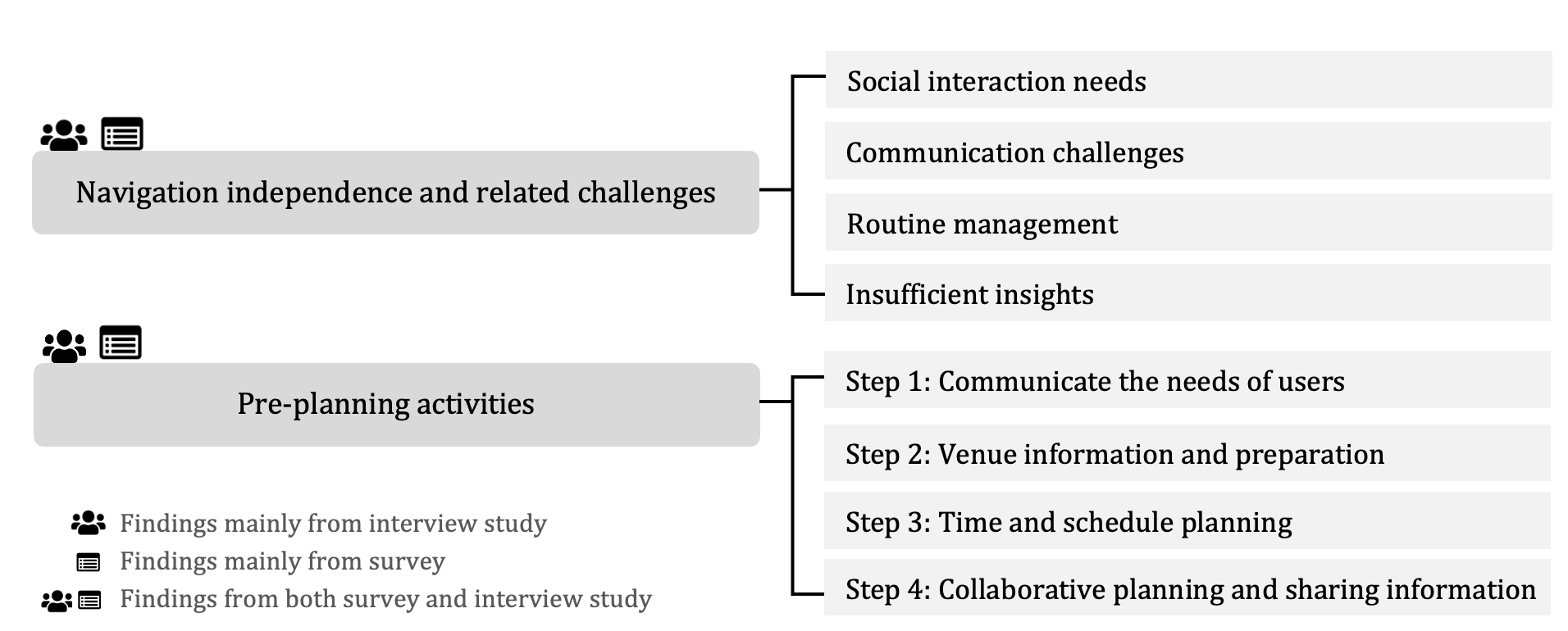}
    \caption{Navigation patterns of IwCI}
    \label{fig:pattern}
\end{figure}

Our data analysis discovered that IwCI possesses distinctive navigation traits and study participants require specific pre-planning steps when they take IwCI on outings, summarized in Figure \ref{fig:pattern}.

\color{darkblue}
\subsubsection{\faListAlt\hspace{0.05cm} \faUsers\hspace{0.05cm} Navigation Independence and Related Challenges for IwCI}

\begin{figure}
    \centering
    \includegraphics[width=0.6\linewidth]{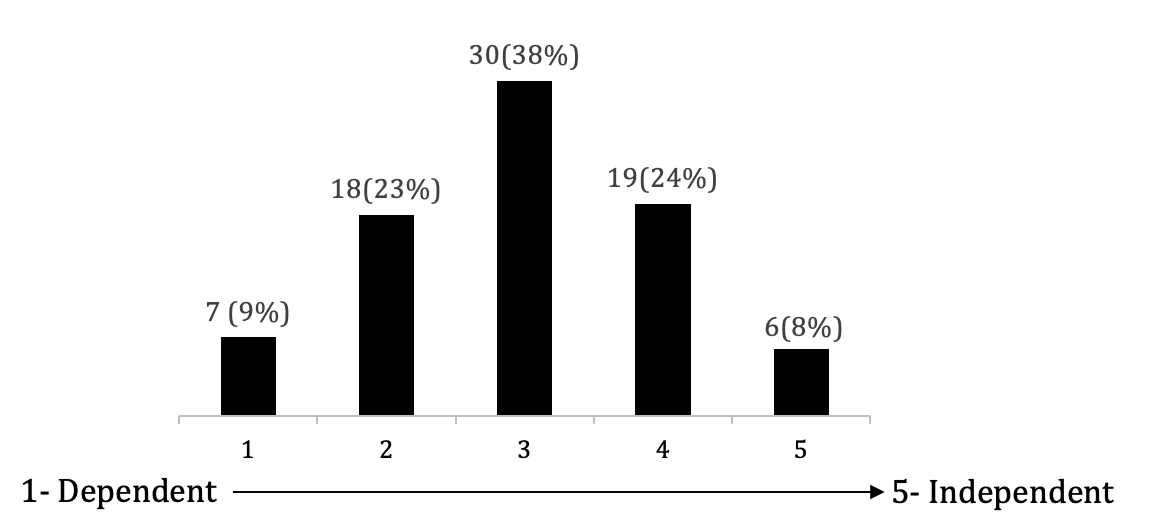}
    \caption{Ability of IwCI to navigate spaces independently from the perspectives of carers}
    \label{fig:dependency}
\end{figure}

In the survey study, participants largely rated the navigation independence of IwCI at level 3 on a scale of 1 to 5, reflecting a moderate degree of independence (See Figure \ref{fig:dependency}). This suggests that while these individuals may navigate independently, they could require occasional assistance or face certain challenges when doing so. Levels 1 and 5, which represent complete dependence or independence, respectively, are the least reported (9\% and 8\%). This suggests that instances at the extremes—where individuals either require complete assistance to navigate or can navigate entirely independently—are not common within this group. IN1, IN9, and IN14 all mentioned that users exhibit a wide range of dependency levels when traveling.

Interviews elaborated on the specific challenges that impact IwCI’s ability to navigate independently. Participants described several interrelated difficulties involving \textbf{\textit{social interactions}}, \textbf{\textit{communication barriers}}, \textbf{\textit{rigid routines}}, and \textbf{\textit{lack of insight}}—all of which can reduce IwCI’s confidence and safety when traveling.  Participants highlighted the importance of facilitating \textbf{\textit{social interactions}} for IwCI, particularly their preference to stay connected with individuals of similar backgrounds or interests [IN3,5,11]. IN11 highlighted that IwCI often thrives when they can interact and socialize with their peers, indicating a requirement to stay with \textit{“their kind.”} This social preference has been shown to improve quality of life, reduce negative influences on IwCI, and decrease social isolation \cite{kuiper2015social}. In public environments, IwCI may encounter challenges when \textbf{\textit{seeking guidance or communication with others}}, which can greatly affect their proficiency in managing through these spaces successfully [IN10,11,15]. This is common among IwCI \cite{johnson2014communication}, leading to feelings of frustration, isolation, and anxiety, and as a result, IwCI may avoid public interactions, limiting their ability to navigate community resources and public spaces confidently. 

Participants also emphasized the importance of understanding and supporting the \textbf{\textit{routines}} of IwCI [IN1,2,4,6,7,9,11,14]. Several participants noted that some IwCI often have rigid routines, which provide structure and predictability, crucial to their well-being and sense of security [IN2,7,11]. For many IwCI, adhering to a routine is essential, particularly when it involves regular activities such as attending medical appointments or visiting family members [IN1,2,4,6]. Research has shown that people with mild to moderate cognitive impairment are capable of consistently answering questions about their preferences, choices, and participation in decisions about daily living \cite{feinberg2001persons}. By honouring their preferences and supporting their routines, caregivers can improve users' independence and satisfaction with their care. Both IN1 and IN14 highlighted that:

\boxquote{Having the ability to store common routes or frequently visited locations in a navigation application can greatly assist users in independently maintaining their routines.}

Participants highlighted the difficulties that IwCI encounter due to \textbf{\textit{insufficient insights}} to identify optimal times and locations for their activities. Several participants noted that users often struggle to understand when and where the perfect time to go would be, leading to difficulties in planning and performing daily activities [IN3,5,6,9,11]. IwCI may rely heavily on recent information or signals, picking up the last few things they heard or saw, leading to inconsistent and often impractical decision making [IN3,5]. 

\color{black}
\subsubsection{\faUsers \hspace{0.05cm} \faListAlt\hspace{0.05cm} 
 Pre-Planning steps}\label{sec:preplan}
\begin{figure}[h!]
    \centering
    \includegraphics[width=0.6\linewidth]{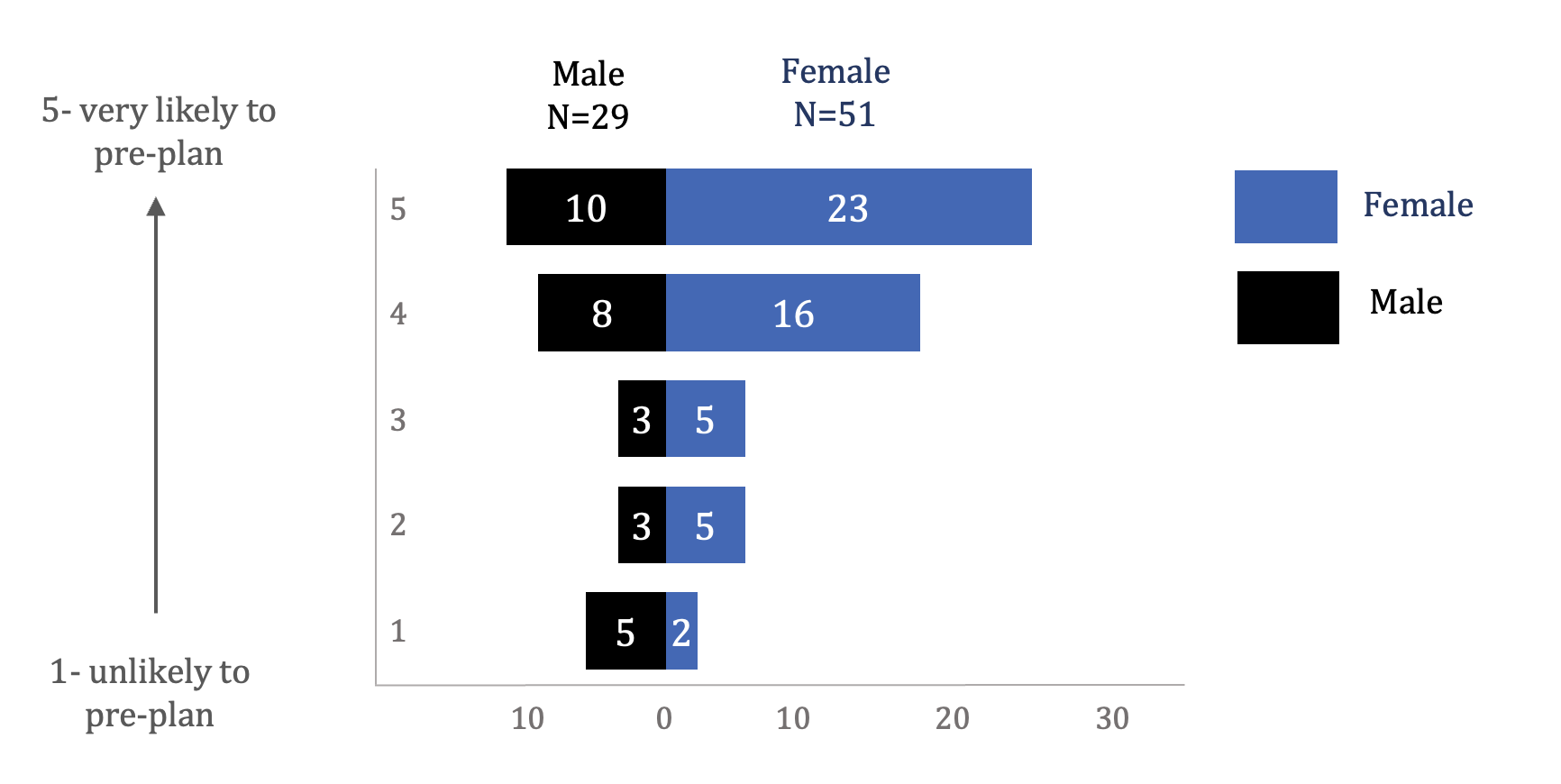}
    \caption{Survey participants' perception of Pre-planning}
    \label{fig:preplan}
\end{figure}

Pre-planning refers to the process of preparing and organizing in advance before undertaking a task or activity. In the context of navigation and support for IwCI, pre-planning involves all steps before the actual time embarking on a journey. In the survey analysis, Figure \ref{fig:preplan} illustrates a noticeable inclination towards pre-planning. Only 9\% of participants are not expected to engage in pre-planning, whereas the tendency progressively rises, reaching 39\% who are very likely to pre-plan. Among the interviewees, a common theme is the expectation that support workers will assist with pre-planning and emphasize the importance of having a plan before traveling. 

\begin{figure}[h!]
    \centering
    \includegraphics[width=0.95\linewidth]{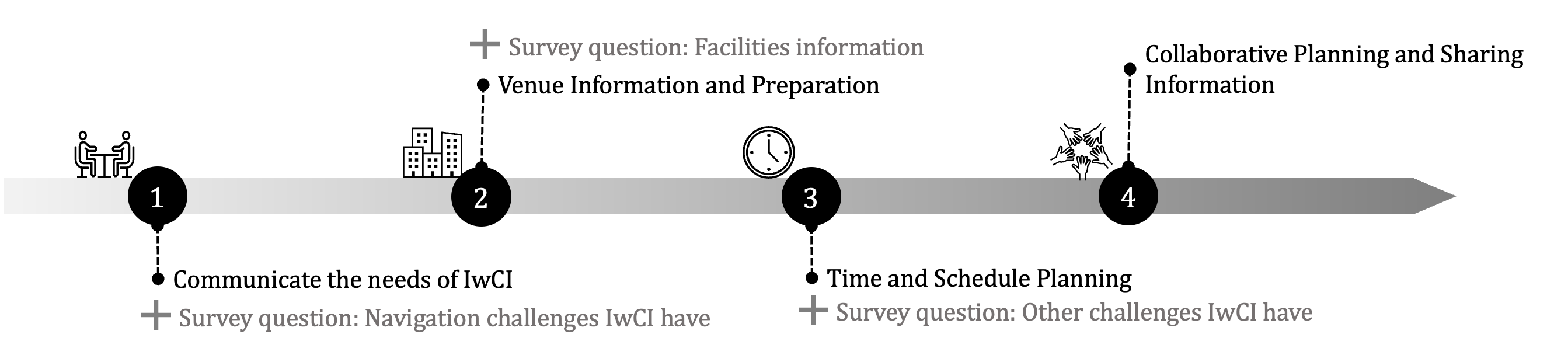}
    \caption{The four steps followed in pre-planning process}
    \label{fig:plansteps}
\end{figure}

The pre-planning part of the navigation experience is often overlooked in the existing literature \cite{yang2018parent,courbois2013wayfinding, davis2014patterns, farran2012useful, farran2016route, purser2015development}. Using \textbf{ thematic analysis} on the survey results, we identified \textit{four steps} of pre-planning undertaken by carers before they go out with IwCI. Specific survey questions focusing on the challenges IwCI faced and the amenities their carers sought provide additional insights on different pre-planning steps, as indicated in Figure \ref{fig:plansteps}.

\textbf{Step 1: Communicate the needs of users.} Both the survey study and the interviews highlighted the importance of engaging with IwCI to understand their challenges and requirements. In the survey study, participants were asked about sensory difficulties and other challenges that IwCI may face. We found auditory difficulties, visual sensory challenges, and motion-related difficulties to be the most common challenges for IwCI (See Table \ref{tab:sensory_difficulties}). The survey participants described how sensory challenges affect IwCI: 

\surveyquote{Experiencing all sensory stimuli together might trigger memories of overwhelming past situations, causing hesitancy to enter specific public areas, such as hospitals.}
Numerous non-sensory obstacles, such as crowds, communication difficulties, challenges in decision-making, and issues with memory, can all hinder the ability of IwCI to move independently through spaces (See Table \ref{tab:sensory_difficulties}).

\begin{table}[h]
    \centering
    \resizebox{0.98\textwidth}{!}{%
    \begin{tabular}{>{\raggedright\arraybackslash}m{16cm}cc}
        \hline
        \textbf{Navigation Needs} & \textbf{Participants} & \textbf{\#} \\ \hline
         \textbf{Sensory Challenges} &  & \\ \hline
        Auditory (e.g., Shopping centres with crowded areas, music.) & 73 & 91\% \\ 
        Visuals (e.g., Overly bright lighting or sudden changes in lighting conditions.) & 63 & 79\% \\ 
        Motion (e.g., Difficulty with spaces that induce motion, such as revolving doors or moving walkways.) & 57 & 71\% \\ 
        Smell (e.g., Strong odours from perfumes or exposure to unpleasant smells in hospitals or public spaces.) & 28 & 35\% \\
        Temperature (e.g., Inconsistent heating or cooling.) & 22 & 28\% \\ 
        Other & 5 & 6\% \\ \hline
        \textbf{Other Challenges} &  & \\ \hline
        Crowds challenges & 69 & 86\% \\ 
        Communication challenges (e.g., communication and social cues, potentially leading to misunderstandings.) & 64 & 80\% \\ 
        Difficulty with decision-making & 61 & 76\% \\ 
        Memory challenges (e.g., remembering directions or locations.) & 50 & 63\% \\ 
        Time challenges (e.g., challenges in estimating time and adhering to schedules.) & 44 & 55\% \\
        Map-reading challenges & 40 & 50\% \\ \hline
        \textbf{Facilities Information} &  & \\ \hline
        Quiet areas & 67 & 84\% \\
        Location of accessible toilets & 60 & 75\% \\ 
        Location of accessible parking & 45 & 56\% \\ 
        Location of accessible seating areas & 43 & 54\% \\ 
        Nearby transportation hubs & 42 & 53\% \\ 
        Location of lifts & 40 & 50\% \\ 
        Nearby cafes or food courts & 37 & 46\% \\ 
        Location of staircase locations & 31 & 39\% \\ \hline
    \end{tabular}}
    \caption{Survey Study: Navigation challenges and Participant Distribution}
    \label{tab:sensory_difficulties}
\end{table}
% Requires: \usepackage{graphicx}

In our interview study, participants also recognized sensory challenges and sensitivities as crucial for avoiding situations that could cause discomfort or distress [IN1,7,8-11,13-15]. They also commented on other information from IwCI. Firstly, it is essential to be aware of the users’ preferences for certain shops or locations during navigation [IN1,7,9,15]. Knowing IwCI’s preferred train lines or travel routes helps in providing smoother and less stressful travel experiences [IN8,9,14,15]. Support workers should be aware of users’ music preferences, which can help create an enjoyable atmosphere during travel or while waiting, particularly when traveling by private car [IN7]. Moreover, understanding dietary preferences or favorite restaurants is useful for meal planning and ensuring pleasant dining experiences [IN1,7,13,14]. IN14 noted that:

\boxquote{Knowing the precise location and kind of food they enjoy is important. For example, my client has a preferred smoothie shop, and it has to be that exact one, not any other similar smoothie shop. Additionally, they only want strawberry smoothies among all options.} 

Keeping informed of the medication schedules of the users is essential for prompt administration and avoidance of health complications [IN9-11,14,15]. This knowledge of IwCI significantly aids in planning routes prioritizing familiarity, as familiar landmarks play a critical role in human wayfinding \cite{hamburger2014role}. Research highlights that environmental familiarity is a key factor that influences behavior in navigation, providing individuals with a sense of comfort and confidence when navigating \cite{meneghetti2017contribution,muffato2020knowledge}.

\textbf{Step 2: Venue information and preparation.} Survey respondents stated that detailed venue information and preparation are sought, including accessibility features, layout and sensory conditions such as lighting and noise, in addition to logistical aspects such as waiting times and available facilities. Personal recognition, consultations with others, and reviews are also recommended for first-hand insights. Survey participants also highlight the facilities they intend to explore during their planning phase (See Table \ref{tab:sensory_difficulties}). Key priorities include quiet areas (valued by 84\% of respondents) and accessible toilets (important to 75\%), emphasizing the need for sensory respite and essential facilities for individuals with disabilities. Accessible parking and seating areas are significant for 56\% and 54\%, respectively, indicating the necessity for mobility and comfort. Additionally, nearby transportation hubs and lifts are valued by a slight majority (53\% and 50\%).

\textbf{Step 3: Time and schedule planning.} Survey participants emphasized the significance of integrating factors such as transportation options, activities, and duration of their stay, using various modes of transport. The interview participant also highlighted the importance of timing in the planning process. Knowing the time required to walk between different locations helps plan efficient and manageable travel schedules [IN9]. Additionally, aligning the travel plan with the support worker's working hours is crucial to ensure continuous and uninterrupted assistance. It would be beneficial to have a reminder that indicates whether the journey to and from a location will exceed the usual working hours [IN8,9,13,14]. In addition, awareness of weather conditions is essential as it can significantly affect travel plans and the comfort of the individuals being assisted [IN7-11,14]. IN14 mentioned specific weather guidelines in place and \textit{"clients cannot be taken outdoors when temperatures exceed certain thresholds"}. Negative weather conditions significantly influence clients' moods, making them fatigued more rapidly.

Participants in the survey are advised to list possible events (such as crowded venues or feelings of anxiety) and to engage in emergency planning, devising strategies to address potential issues such as anxiety for IwCI in crowded areas. IN4 and IN6 also suggested having backup travel plans in case the main transportation method is unavailable. This includes being aware of the conditions of the roads and pathways to avoid possible barriers and guaranty a safe journey [IN4]. Previous research has recognized that barriers such as steep inclines, stairs, rocks, and uneven ground along chosen paths significantly impact the navigation experience, particularly for individuals with mobility impairments \cite{harriehausen2016communicating}. IN4 specifically noted that:

\boxquote{ I am familiar with the roads in the community when I accompany clients. I prefer routes that are smoother and have fewer traffic lights. Navigating inclines is particularly challenging for clients with mobility issues, and some traffic lights have very short duration, making it difficult to cross in time. Occasionally, we are still in the middle of crossing when the light turns red.}

\textbf{Step 4: Collaborative planning and sharing information.} Survey participants finally highlighted that collaborative planning underpins the process, involving clear communication with IwCI and support networks to discuss potential difficulties and offering provisional plans to address feelings of being overwhelmed, ensuring a comprehensive and inclusive pre-planning process. The interviewees also highlighted the importance of collaborative planning, with further details available in Section \ref{sec:support}.

\subsection{Improved navigation software design}\label{interview}
We wanted to gather some requirements and high-level design principles from our study participants that could lead to practical improvements in navigation software for IwCI. In our survey and interview studies, we discovered that both IwCI and their carers expect additional assistance from the navigation software they use. In particular, we identified unique support needs among three different groups of stakeholders, summarised in Figure \ref{fig:All}.
%---------------------First category-------------------%

\subsubsection{\faUsers \hspace{0.05cm} \faListAlt\hspace{0.05cm} Navigation support in navigation software for IwCI}\label{sec:support}
\begin{figure}[h!]
    \centering
    \includegraphics[width=0.95\linewidth]{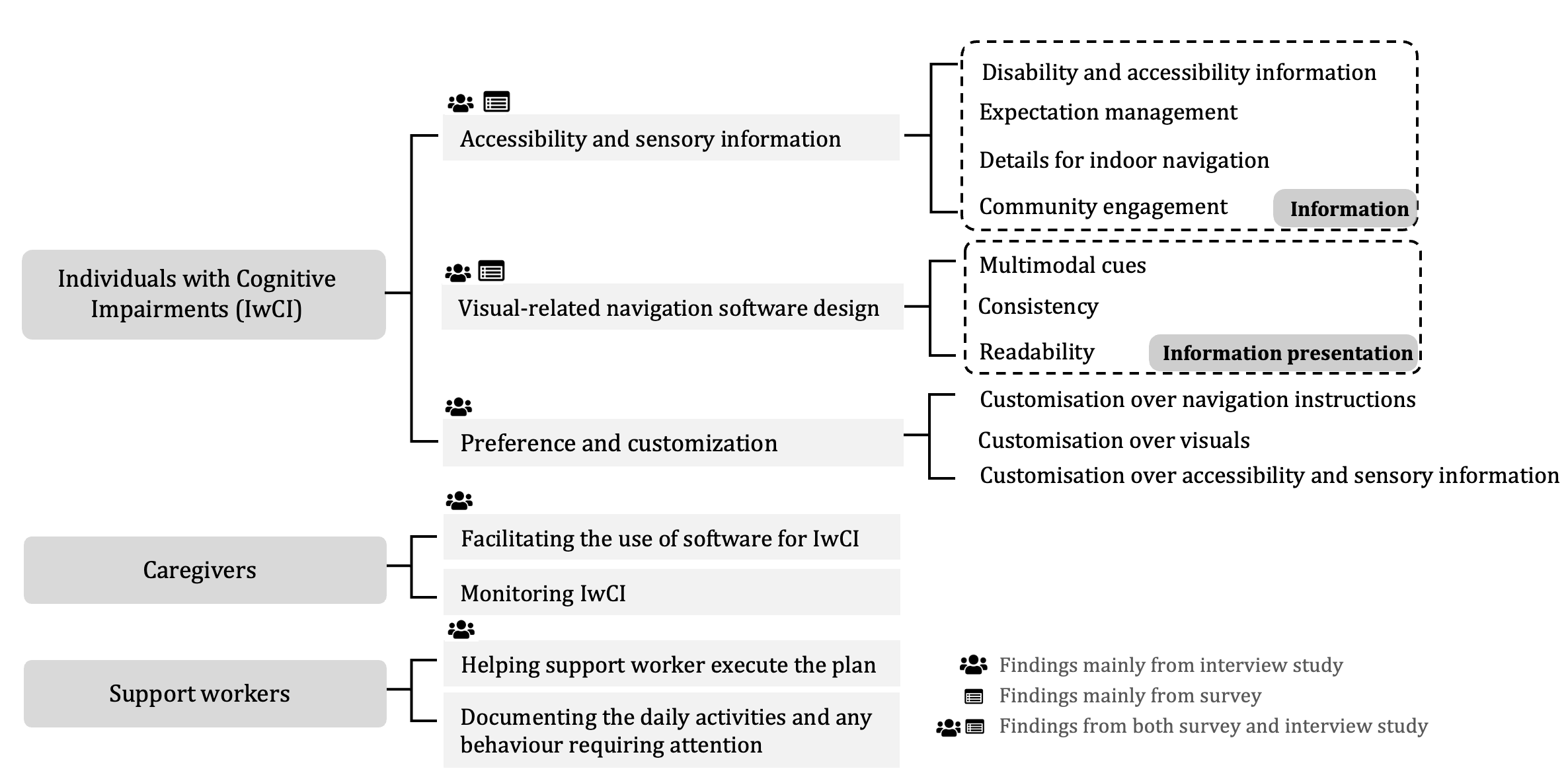}
    \caption{Navigation support for IwCI, caregivers and support workers}
    \label{fig:All}
\end{figure}

The primary source of navigation assistance required for IwCI is mainly from healthcare professionals, caregivers, and support workers, as revealed in the interview study. Navigation support is divided into three primary subcategories that encompass the broader concepts of \textit {information, presentation of information, and customisation} (See Figure \ref{fig:All}). This suggests that the system must deliver essential data concisely and clearly while tailoring itself to meet the specific requirements of each user.

\textbf{Accessibility and sensory information.} This category entails integrating various accessibility and sensory details into navigation software, which are crucial for IwCI who encounter different levels of accessibility issues and sensory sensitivities. Support workers, health professionals, and caregivers focus primarily on determining if a location is suitable for individuals with various \textit{disabilities or sensory challenges}. All interviewees highlighted the need to have information on accessibility features for users with disabilities, including the availability of disabled toilets. In addition, most of them mentioned the importance of various sensory details for the places they intend to visit on the map, such as temperature, noise, crowd density, smell, and lighting. This is a critical factor because spatial navigation is a multi-sensory process that requires integrating and manipulating information over time and space \cite{wolbers2010determines}. Existing research examining sensory challenges is primarily concerned with designing navigation tools for blind users, focusing on choosing their preferred modes to provide information and communication \cite{gallagher2014indoor,rodriguez2014accessible, dunai2013sensory}. Some interviewees also highlighted the importance of being aware of the intensity levels of various sensory inputs [IN3,5,7,10,11]. For example, IN3 added that:

\boxquote{ I usually go to the Coles[supermarket] or Woolworth[supermarket] by checking the busy level of the supermarket. Likewise, for users with cognitive impairment, if the level of different sensory information can be seen on the map so their caregivers can view and maybe decide when is the best time for them to go to that place.} 
Accessibility characteristics were also particularly highlighted by the survey participants, with the presence of disabled-friendly amenities, clear signage of lifts and wheelchair access routes, and the inclusion of digital maps showing key facilities such as toilets and information desks. QR codes were noted to provide quick access to information. Several survey participants noted:

\surveyquote{Maps not showing accurate accessibility features like ramps and providing outdated or inaccurate information is very frustrating. This often leads to unclear directions and inaccuracies in location details}
\indent Several participants highlighted the necessity of setting clear \textit{expectations} regarding the users' current position and the distance they need to cover, whether walking or driving[IN4,5,6,9,10,12,14]. For example, some participants think that implementing a \textit{"navigation progress bar"}, which includes distance milestones, would be advantageous for IwCI [IN4,5,6]. Previous studies have developed methods that allow users to preview upcoming steps and review previous instructions with a single click or voice command, thus giving IwCI more control and autonomy while using the application \cite{gomez2015adapted}. Furthermore, some interviewees suggested displaying the current location of the user with an image showing the surroundings, such as what is left and right on the map, to boost their confidence and recommended highlighting familiar landmarks as users pass them [IN4,9,10,12,14]. Previous studies have explored the incorporation of street-level landmark images with instructions to assist IwCI in identifying specific locations more easily\cite{gomez2015adapted, liu2009customizing, lewis2015hearing}. Survey participants also commented on the real-time \textit{"You are here"} markers and functionalities that assist in identifying one's present location and planning routes, incorporating vertical navigation across various building levels.

The details of a location are also crucial for healthcare providers, support workers, and caregivers. They anticipate locating the quiet room, exits, entrances, elevators, and accessible restrooms. Moreover, they emphasize the possibility of \textit{outdated }online information, leading them to consistently examine user reviews for actual accessibility and sensory information. For instance, IN8 recounts an incident where the elevator was too small to accommodate a wheelchair for an individual, and certain office spaces were also occasionally quite cramped. This forces them to look for detailed and precise information about the specific location they intend to visit. The survey participants also noted that the ability to \textit{engage with community} input through comments and reviews was perceived to enhance the overall utility of the navigation system. Almost all support workers indicated that they seek more up-to-date information online. IN9 highlighted this aspect.

\boxquote{I find online information unreliable because it can be outdated or incorrect due to ongoing construction. To ensure accurate details, I prefer to consult recent visitors. I usually check Google Maps reviews, talk to people who have been there, or contact the venue's staff before visiting. }

\textbf{Visual-related navigation software design.} Participants indicated a need for the use of different \textit{multimodal cues}, incorporating text and visual components, such as real street photos or cartoon drawings, to assist users who prefer visual interpretation [IN1,7,9,12,14]. Several participants underlined the need to use multiple methods to present information, as visuals alone may not always be comprehensible to all users [IN1,7,12]. This was also highlighted in an earlier study that focused on multimodal interaction with older adults\cite{jian2013touch}. The visuals must be created to make them easy for a child to understand [IN1] and faithfully depict the actual landmarks or settings experienced by the users [IN1,12].

Concerns about \textit{consistency} frequently emerged among the participants, both within navigation apps and in their correlation with the real world. Within the app, maintaining a consistent design through the use of colour coding for specific places helps to ensure that users can more easily navigate and recognize different areas, creating a predictable and familiar interface [IN1, 6,7,11,12,15]. Previous research has investigated the use of consistent color schemes to help visually impaired users identify specific obstacles while navigating various environments\cite{als2018blukane}. Furthermore, research on audio-guided systems for seniors has emphasized the use of consistent sound signals to differentiate between various scenarios \cite{jian2013touch}. In addition, the icons and symbols in the app should closely resemble their real-world counterparts, ensuring that visual representations have clear and understandable links to actual objects or places [IN1,5-9,12]. Our survey respondents indicated a preference for modifications in color (56\%) and icons (50\%). They also emphasized the importance of alignment between map colors/floors and the actual buildings to facilitate easier correspondence between map navigation and physical structures. This alignment between the app and reality helps IwCI intuitively connect the interface with the real world, facilitating easier navigation and reducing cognitive load. 

Maintaining \textit{readability} is crucial for effective user interaction. This can be achieved by providing sufficient contrast for clarity and highlighting changes. Participants highly recommended that navigation should include sufficient contrast to improve clarity, better distinguish text and elements, and mitigate eye strain [IN1,3,5,13,14]. The survey participants mentioned the requirement of showing important locations in bold colors. They suggested using the categories of the places to differentiate the colors while ensuring that the hues are distinct from one another (for instance, avoiding colors that are too close, like yellow and gold). This is particularly important for users with visual impairments and has been the subject of extensive research in design efforts to help visually impaired individuals with navigation\cite{jian2013touch,griffin2020effectiveness,als2018blukane}. Furthermore, key features should be accessible without scrolling to allow users to quickly and easily find important information [IN1]. This approach minimizes additional navigation requirements and helps users focus on key content, as highlighted by several studies\cite{krainz2016accessible,jian2013touch}

\textbf{Preferences and customisation.} The requirements of IwCI and their carers when it comes to navigation software are varied, making a one-size-fits-all approach impractical. Our participants stressed the need for users to personalize their navigation experience. This can involve modifying navigation guidance, visual features, and overall accessibility settings to align with personal preferences. All support workers we interviewed consistently emphasized that IwCI differs from each other. IN11 highlighted this point:

\boxquote{It's hard to determine their needs because each individual is unique. Therefore, we will talk with our clients and their caregivers to understand them and always communicate with them to know what they want and what they are struggling with. You can only apply the same rules to all clients}

\textit{Customisation of navigation instructions} is crucial to address the varied requirements and preferences of IwCI. IwCI should be able to modify the interaction settings for their comfort [IN4,5,6,8,13], including various language options for those who are not native speakers [IN7,9,14], customizable speech instruction speed [IN1,7,14], and the option to choose between metric and imperial units for measuring distance [IN1,3,9,11,14,15]. These aspects have also been explored in several studies focused on visually impaired users, including the integration of audio and vibration feedback to help blind users in noisy settings\cite{rodriguez2014accessible}, as well as the analysis of conversion between imperial and metric units\cite{lewis2015hearing}. Much of this work is mainly focused on understanding and meeting the needs of visually impaired users. Insights and proposals from these studies are sometimes also highlighted by individuals with experience working with and living alongside IwCI. This indicates a transfer of expertise and experience, where the knowledge gained from assisting visually impaired individuals is used to benefit IwCI. There may be overlapping challenges among various disabilities that could be tackled using similar design approaches, and it is also likely that the IwCI mentioned by the participants possess some level of visual impairment. Some support workers have observed that IwCI can face both mobility and vision challenges [IN1,7,9,11,15]. Studies also indicate that older adults often experience visual and cognitive impairments together, and these conditions are interconnected \cite{uhlmann1991visual,lim2020association}.

\textit{Visual customisation} was also highlighted by some of our participants [IN1,3,7,12]. Users should be able to modify visual elements like font size, color schemes, and contrast settings to suit their personal preferences and requirements. IN1 and IN12 specifically emphasized that images can be very impactful but must be suitable, considering the design's style and users' familiarity with the content. Users should be offered a selection of system-supplied visual themes and allowed to upload images used by support workers or caregivers in their daily tasks.

\textit{Customization of accessibility and sensory information} is essential for designing a user-friendly experience that meets individual requirements. Users should be able to prioritize routes according to various accessibility and sensory attributes, ensuring that their specific requirements are addressed during navigation. This involves choosing routes that consider mobility constraints, sensory preferences, and various other accessibility requirements. Furthermore, users should be able to prioritize destinations that best align with their accessibility and sensory needs, ensuring a more personalized and pleasant experience [IN2,4,14,15]. IN14 also highlighted the importance of allowing users to classify locations according to their preferences. The guide facilitates the creation of personalized place categories, allowing users to define categories such as \textit{"Quiet Cafes"}, \textit{"Pet-Friendly Parks"}, or \textit{"Accessible Stores"}, in line with their specific needs. Categories or destinations can be displayed on the smartphone screen using icons and names \cite{gomez2015adapted}. This approach allows the system to present IwCI with a list of potential destinations, allowing them to easily recognize and select one, thus minimizing the cognitive effort required\cite{friedman2007web}.

Customisations are crucial in setting expectations for a location, enabling users to specify their preferences and expectations for particular places. Users can enter details such as the desired ambience, accessibility options, or key points of interest for each location. The guide improves the alignment with users' expectations by customising navigation instructions according to these personalised profiles. Recognising the importance of place expectations, the navigation guide emphasises incorporating features that match users' anticipated experiences in specific locations. 

% Several participants emphasise on the importance on the managing expectation that IwCI and they have to reminder them for what they can expect from the system.

%---------------------Second theme-------------------%
\subsubsection{\faUsers \hspace{0.05cm} Navigation support in navigation software for caregivers}\label{sec:support}
Participants highlighted several key recommendations to support caregivers navigate more efficiently and improve the quality of care they provide. As described in Figure \ref{fig:All}, the key tasks of caregivers include \textbf{\textit{facilitating the use of the app for IwCI}}, as well as \textbf{\textit{monitoring IwCI}} to ensure their safety and comfort. Caregivers emphasize the importance of facilitating the use of the app and monitoring users [IN2,4,6]. These capabilities can improve caregivers' proficiency in managing care for individuals, possibly through a navigation app devised to assist their responsibilities. Firstly, they anticipated the app to provide functionalities that simplify its usage for caregivers [IN2,4,6]. It is essential to integrate features that enable caregivers to support users efficiently. Carers should be able to adjust the user settings as mentioned previously in Section \ref{sec:support} [IN2], ensuring that the application accommodates particular sensory sensitivities and accessibility requirements. Moreover, carers should be able to modify destinations for users, aiding in more efficient route planning and management [IN4,6]. In the study conducted by \citet{gomez2015adapted}, caregivers must select destinations for users on the website portal, which would then be displayed on the user's phone. This process reduced the cognitive load on users and supported the caregivers in their role. These features allow caregivers to offer personalized support, enhancing the app's accessibility and user-friendliness for IwCI. IN2 particularly highlighted that the central focus should \textit{"not merely be on the application's usage or on adding fancy features"}; without supplementary support and functionalities, IwCI could easily become overwhelmed and struggle with decision-making. 

Another important aspect they highlighted, particularly, is the need for caregivers to monitor IwCI to ensure their safety and well-being effectively. Effective communication between caregivers and IwCI should be ensured, allowing caregivers to offer timely help when necessary [IN2,4,6]. For scenarios where IwCI deviates from their intended path or is delayed, caregivers must be notified to offer guidance or reassurance [IN6]. Participants stressed that caregivers should have access to identical maps, allowing them to monitor users' locations and routes precisely, and provide updates on users' whereabouts and expected arrival times [IN2,4]. This recommendation raises ethical concerns, as individuals should not be forced to use tracking technology; whenever feasible, IwCI should participate in the decision-making process and their consent should be obtained. Research has explored the use of tracking technology for people with dementia \cite{landau2012ethical,oderud2015persons}. It has been recommended that decisions about timing and tracking methods be made with the input of the person with dementia, their family, and professional caregivers within formal structured meetings led by a professional team. This indicates that the choice to implement tracking for IwCI should be a collective decision-making effort, including input from experts, families, and users themselves.

\subsubsection{\faUsers \hspace{0.05cm} Navigation support in navigation software for support workers}
Support workers highlighted the need for comprehensive information to effectively manage pre-planning activities (as discussed in Section \ref{sec:preplan}). As described previously in Figure \ref{fig:All}, it was also noted that they anticipated having access to tools that \textbf{\textit{facilitate plan execution}}, and \textit{\textbf{support thorough activity documentation}}. Both IwCI and venue information should be handled efficiently using a feature such as \textit{“keyword search,”} enabling support workers to find relevant details using specific keywords [IN7,8,10]. This feature enables quick access to necessary details such as “wheelchair accessibility,” “quiet zones,” “restrooms,” or “dietary options,” making the planning process more efficient and tailored to the needs of IwCI. Implementing keyword search functionality can streamline the information retrieval process and ensure that support workers have immediate access to pertinent details when needed. 

Support workers have different needs regarding the execution of the plan, highlighting the importance of offering practical tools to successfully implement care plans. 
Several participants highlighted the need for features that help in the setup of typical user profiles to facilitate further searches [IN8,9,11,14]. This includes creating detailed profiles that describe user preferences, routines and specific needs, which can streamline the planning process. IN9 also noted that the ease of printing the user profile would also be of great help. 
A step-by-step guide for the day’s plan is crucial to provide clear instructions and ensure consistency of care [IN9,10,14]. In addition, tools that facilitate collaboration between support workers, caregivers, and users are essential for efficient and coordinated care [IN7,10,11-14,15]. One suggestion is the ability to email the plan to caregivers, ensuring that everyone involved is informed and on the same page [IN11]. 

Communicating the time and duration of each part of the plan is also important to manage expectations and maintain a structured schedule [IN13,15]. IN15 specially noted factors such as weather can significantly influence these experiences, potentially shortening or extending the perceived duration of activities. Visual communication aids, such as storybooks, can be used to explain the plan to users, making it easier for them to understand and follow [IN12,13]. Providing an alliterative routine or activities that can be used when users are overwhelmed, such as the fastest or easiest routes, helps to manage emergencies and reduce stress for both IwCI and support workers [IN9,10,13,15]. IN15 specifically mentioned the unpredictable nature of some places and weather and noted that using the navigation app's point-of-interest feature to mark indoor venues along the route makes it easy for support workers to navigate to these locations. Several participants noted that due to COVID-19, most documentation and pre-planning services have transitioned online. They use a dedicated app or website to access various types of information. However, they find it inconvenient to pull out their phone and navigate to a specific page while traveling with IwCI. They expressed a desire to print out the plan or essential user information [IN9,11,14]. IN9 emphasized that:

\boxquote{We conduct numerous activities with client groups. Frequently, I join a group of clients which might include three people. I rarely find the time to check my phone, and it's difficult to interact with clients through my phone if they are uncertain about something and I need to clarify it. Printed paper allows clients to record locations or mark specific areas if they wish.}

Participants also emphasized the importance of comprehensive post-documentation to safeguard client welfare and adhere to funding regulations. Several participants [IN2,5,7,11] highlighted the need to maintain complete documentation, including daily notes, incident and security notes, client and family communication notes, change and support notes, and result notes. Daily notes provide a detailed account of user activities, helping to monitor their routine and identify any deviations that may require attention [IN3]. Incident and security notes are crucial to documenting unusual or emergency situations, ensuring that all incidents are recorded for future reference and action [IN8]. Client and family communication notes help maintain a clear line of communication between support workers and the family, ensuring that everyone is informed and involved in the care process [IN6]. Change and support notes are essential for recording changes in the user’s condition or care plan, ensuring that all adjustments are tracked and communicated effectively [IN10]. Outcome notes document the results of the care provided, helping to assess the effectiveness of the interventions and identify areas for improvement [IN9].
%---------------------Third category-------------------%

%------------------Discussion----------------------%
\section{Discussion}
\begin{figure}[h!]
    \centering
    \includegraphics[width=0.97\linewidth]{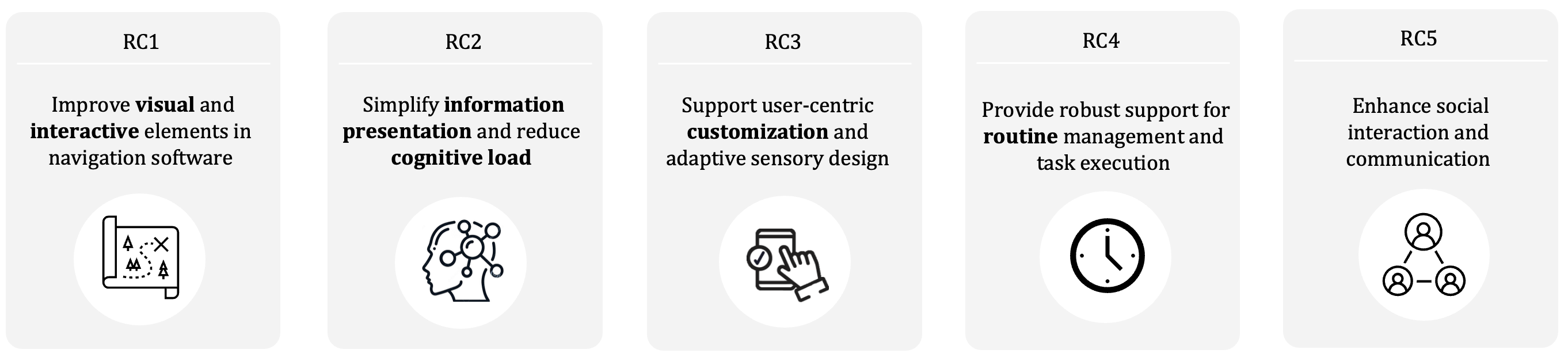}
    \caption{Design recommendations for navigation software for IwCI}
    \label{fig:recomIwCI}
\end{figure}

\begin{figure}[h!]
    \centering
    \includegraphics[width=0.99\linewidth]{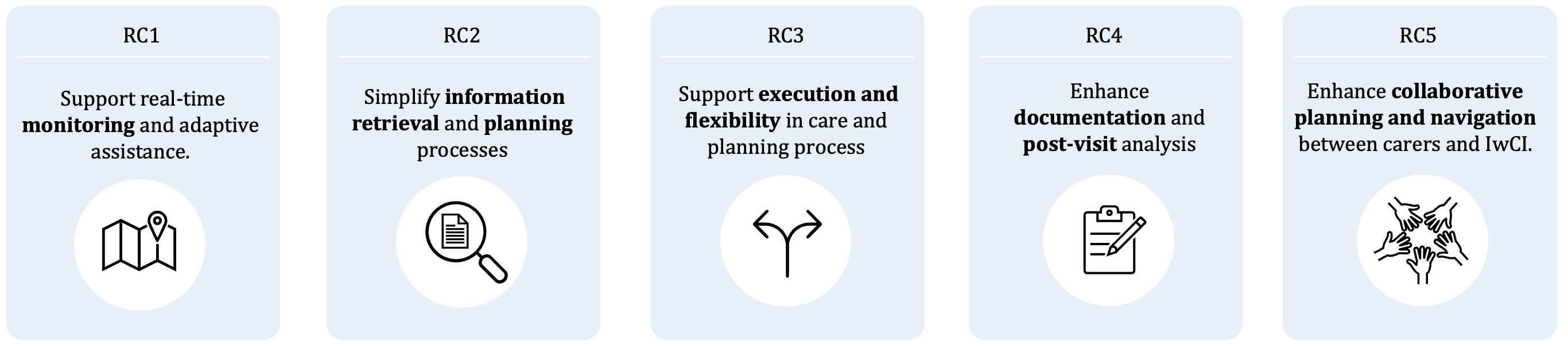}
    \caption{Design recommendations for navigation software for caregivers and support workers}
    \label{fig:recomcarers}
\end{figure}

\color{darkgreen} This section outlines key design recommendations for developing inclusive navigation systems that support IwCI and their carers (Figures \ref{fig:recomIwCI} and \ref{fig:recomcarers}). These recommendations are grounded in empirical data collected across both indoor and outdoor navigation contexts, including environments such as public transport stations, medical centres, shopping malls, and pedestrian streets. By capturing this diversity of navigational settings, the recommendations address a broad spectrum of spatial, sensory, and social challenges encountered by IwCI in everyday life. At a high level, these recommendations extend current RE practices by illustrating how stakeholder-informed, empirically grounded design requirements can inform inclusive software specifications. Both survey and interview responses strongly emphasized that individuals with cognitive impairments differ significantly in their support needs, preferences, and ways of processing information. To interpret our recommendations for IwCI specifically, it is important to recognize that customization and configurability are embedded as an underlying principle across all recommendations. Rather than prescribing a fixed design for all users, our recommendations promote flexible interaction models that can adapt to the diverse cognitive profiles within the IwCI population. While grounded in empirical data, our recommendations are intended as cross-cutting design principles rather than platform-specific prescriptions. They are applicable to a variety of navigation technologies—including mobile apps, VR systems, and assistive wearables—that support IwCI and their caregivers. Translating these principles into actionable features will require attention to the affordances, constraints, and usage contexts of each platform. For example, real-time monitoring may be operationalized differently across smartphone, smartwatch, or immersive VR settings.

\subsection{Design recommendations for navigation software for IwCI}

\color{darkblue}
\indent\hspace{0.1cm} \textbf{RC1: Improve visual and interactive elements in navigation software:} Improving visual and interactive elements is essential to reducing cognitive load and improving usability for IwCI. To achieve this, navigation interfaces should prioritize visual clarity and consistency, incorporating high-contrast colors, large text, and easily recognizable icons to support users with perceptual or attention-related challenges. Furthermore, maintaining alignment between digital interfaces and real-world signage systems enhances user orientation and continuity between virtual and physical spaces, an important aspect of cognitive mapping. To foster engagement and comprehension, navigation tools should include interactive and multi-sensory features—such as 3D map views, street-level imagery, and community-generated annotations (e.g., reviews, accessibility comments). These interactive affordances not only improve situational understanding but also empower users to anticipate environmental conditions, reducing anxiety during navigation. This recommendation is supported by 56\% of survey respondents who identified visual clarity as a critical need, and 81\% of interviewees who emphasized the importance of consistent visual elements. Our findings align with the existing literature on navigation systems for individuals with cognitive impairments, particularly in the areas of visual impairment \cite{somyat2018navtu, ganz2014percept, prerana2019stavi, paladugu2013supporting}. For example, previous studies have highlighted the importance of consistent visual elements in reducing cognitive load \cite{krainz2016accessible,jian2013touch}.

\hspace{0.1cm} \textbf{RC2: Simplify information presentation and reduce cognitive load:} 
Interfaces should prioritise clarity, minimalism, and information hierarchy, presenting only the most essential details while allowing users to selectively expand or filter supplementary information. This approach aligns with \textit{progressive disclosure principles} in software design, ensuring that users are not overwhelmed by excessive data during decision-making \cite{springer2019progressive}. To support orientation and recognition, the system should incorporate consistent visual cues, such as colour coding for specific places or landmarks, enabling users to intuitively associate distinct areas with specific colours or icons. This approach promotes a predictable and familiar environment, which has been shown to improve spatial memory and reduce navigation anxiety \cite{shove2007design}. In addition, implementing progress indicators—for example, route progress bars that show current location, remaining distance, and upcoming milestones—can help users maintain situational awareness throughout their journey. These indicators also improve confidence by providing clear sequential signals, an effect supported by previous research that highlights the role of structured pathfinding information in maintaining orientation \cite{passini1996wayfinding,lidwell2010universal}.

Empirically, this recommendation is supported by 63\% of survey respondents and nearly all interviewees, who emphasized clarity and simplicity as core aspects of effective navigation tools. From a requirements engineering perspective, RC2 underscores the need to treat cognitive load management, guiding software engineers to specify constraints on information density, interaction pacing, and visual complexity. This integration ensures that accessibility and comprehension are built into early design stages rather than retrofitted during implementation.

\hspace{0.1cm} \textbf{RC3: Support user-centric customization and adaptive sensory design.} Empowering IWCI through personalized and adaptive interfaces is essential for inclusive navigation systems. Navigation tools should provide user-centric customization options that allow IWCI to adjust visual elements such as font size, color contrast, and map icon clarity. Beyond visual customization, personalization of navigation instructions—including language choice, simplified route descriptions, and preferred measurement units (e.g., metric or imperial)—enhances accessibility and user control. To address sensory sensitivities, systems should integrate live environmental data (e.g., ambient noise levels, crowd density, lighting conditions) to allow users to select or avoid certain routes based on their comfort preferences. This design reduces overstimulation, improves perceived safety, and fosters confidence when navigating complex public spaces, strengthening findings from previous studies on sensory-inclusive design \cite{williams2014creating}. Moreover, tailoring the complexity of information—for instance, by dynamically simplifying route details or hiding extraneous interface elements—enables the system to adapt to varying cognitive capacities and situational demands. Empirical evidence from this study reinforces these principles: 12 out of 15 interview participants explicitly emphasized the need for customization to accommodate the diverse preferences and abilities of individuals with cognitive impairments. Existing research has largely overlooked such adaptive mechanisms, often designing for single-condition user groups rather than acknowledging variability across cognitive profiles \cite{liu2006indoor, gomez2015design, chang2010autonomous}.

\hspace{0.1cm} \textbf{RC4: Provide robust support for routine management and task execution.} Routine stability is a cornerstone of independence for individuals with cognitive impairments, making routine-oriented design a critical requirement for inclusive navigation software. Prior research underscores the importance of maintaining predictable and structured environments to reduce anxiety and cognitive effort during daily activities \cite{woo2020routinoscope}. To operationalize this, navigation systems should incorporate routine management features that allow users to store, retrieve, and reuse frequently visited destinations and preferred routes. Automating these repetitive tasks reduces input effort, improves efficiency, and fosters user familiarity with the interface—key enablers of sustained engagement. 

In parallel, systems should provide adaptive support for disruptions by integrating scenario-based guidance capable of dynamically offering alternative routes or schedules in response to contextual changes such as road closures, construction, or crowd congestion. This feature enhances both resilience and continuity, enabling users to maintain autonomy even in unpredictable situations. Anticipating potential disruptions and proactively suggesting viable alternatives increases user confidence and trust in the system, which are essential indicators of long-term technology acceptance. Empirical findings from our interviews indicate that approximately 50\% of respondents reported that IWCI adheres closely to structured routines.

\hspace{0.1cm} \textbf{RC5: Enhance social interaction and communication.} Effective social interaction is a critical but often overlooked dimension of inclusive navigation design, particularly for IWCI. Social interaction plays a vital role for IwCI as it supports emotional well-being, fosters a sense of belonging, and reinforces independence by enabling them to navigate, communicate, and participate more confidently in everyday social environments \cite{bottema2017glimpses,kuiper2015social}. Beyond spatial guidance, navigation systems can serve as social mediators, supporting users to engage with others and managing communication-related challenges in public contexts. Integrating assistive communication features—such as simplified language interfaces, visual conversation prompts, and context-sensitive communication aids—can help users better understand, interpret, and respond to social cues. These features enable smoother interactions in everyday situations such as asking for directions, navigating service counters, or coordinating with caregivers. Several interviewees underscored that social withdrawal is a common consequence of navigation-related anxiety, highlighting the importance of designing for both functional inclusion (e.g., wayfinding support) and social inclusion (e.g., confidence in interaction). RC5 expands the inclusivity agenda from individual usability toward sociotechnical integration, where systems are designed to mediate not only between users and tasks but also between users and their social environments. This perspective situates communication features as relational requirements that bridge human, technological, and social dimensions.

\subsection{Design recommendations for navigation software for caregivers and support workers}
\hspace{0.1cm} \textbf{RC1: Support real-time monitoring and adaptive assistance.} Real-time monitoring is essential to ensure the safety, autonomy, and well-being of IwCI. Implementing caregiver monitoring features that allow caregivers to track the real-time location and progress of users provides an added layer of security. These features can include alerts for deviations from planned routes or unexpected delays, enabling caregivers to respond promptly to potential issues. The interview data strongly supported this need, and all three caregivers highlighted the importance of responsive location awareness—for example, IN4 noted: \textit{“Receiving real-time updates on the user’s location and their progress helps me feel confident that I can intervene if necessary.”} However, such monitoring raises important ethical and privacy concerns that must be explicitly addressed in system requirements \cite{landau2012ethical, oderud2015persons}. It is essential to ensure transparency in tracking features. Providing users with clear options to consent to or decline monitoring fosters trust and respects their independence. In addition, offering customizable privacy settings allows users to control the timing and manner of location sharing, ensuring that they feel comfortable with the level of monitoring in place. To further enhance safety, introducing automated regular check-ins during journeys allows users to confirm their well-being or request help when needed. Embedding these ethical safeguards within early RE activities (e.g., through stakeholder negotiation and scenario-based elicitation) ensures that autonomy and protection are not competing priorities, but rather coexisting design goals within inclusive navigation systems.

\hspace{0.1cm} \textbf{RC2: Simplify information retrieval and planning processes.} Efficient information retrieval and planning are essential for ensuring the usability and effectiveness of navigation systems designed for IwCI and their support networks.Integrating keyword search allows IWCI, their caregivers, and support workers to quickly access relevant information, such as "wheelchair accessibility" or "quiet zones," ensuring that planning is quick and precise. The inclusion of user profile–based personalization allows navigation systems to automatically prioritize and recommend options aligned with individual needs, reducing cognitive effort and decision fatigue. Personalized pre-planning tools can allow caregivers to label and prioritize essential amenities such as quiet areas, accessible facilities, and sensory-friendly zones, catering specifically to the unique needs of ICWI. Empirical findings from our study reinforce this need: 91\% of survey participants reported engaging in pre-planning when accompanying IwCI, and multiple interviewees emphasized the importance of accurate and accessible planning tools to manage variability in real-world conditions. In addition, navigation tools should provide up-to-date and detailed information on accessibility features, sensory conditions, and logistical aspects of venues, empowering caregivers and users to make informed decisions. Embedding adaptive information retrieval mechanisms and context-aware updates in early requirements elicitation phases strengthens user trust, enabling navigation systems to evolve from static tools into dynamic, user-centered decision-support systems.

\hspace{0.1cm} \textbf{RC3: Support execution and flexibility in care and planning processes.} Execution tools play a critical role in bridging the gap between planning and real-world action, particularly for IwCI and their caregivers. These guides, coupled with printable user profiles and storybook-style communication aids, simplify complex tasks and ensure all critical information is easily accessible. By streamlining the execution of plans, these tools also foster better communication and understanding, making the process more intuitive and less stressful. The value of flexible planning cannot be overstated—features like alternative route suggestions or stops in response to disruptions, such as construction or crowded areas, help minimise stress during outings. Additionally, syncing pre-planned routes and essential details with calendar applications offers timely reminders, enabling users and caregivers to prepare in advance. This recommendation is endorsed by the majority of support workers among interview participants, who highlighted the crucial role of assistance in implementing the plan. Flexibility and support in execution are conceptualised here as behavioural quality attributes, promoting dependability, context awareness, and continuity of care. 

\hspace{0.1cm} \textbf{RC4: Enhance documentation and post-visit analysis.} Post-visit documentation tools are vital for improving continuity of care and supporting informed decision-making in future planning. Structured digital interfaces enable caregivers to record essential information like daily notes and reports systematically, ensuring easy retrieval. Beyond documentation, leveraging this accumulated data to generate actionable insights can improve care quality. By analyzing patterns, such as identifying optimal times for activities or predicting potential challenges based on previous experience. Our interviews highlight the importance of these features. Many support staff emphasized the need to document daily activities and incidents for better communication and planning. However, they found that current systems are cumbersome and lack analytical features. This recommendation reframes documentation as a data lifecycle requirement, extending beyond simple record-keeping toward knowledge-based system enhancement. 

\hspace{0.1cm} \textbf{RC5: Enhance collaborative planning and navigation between carers and IwCI.} Enhancing collaboration in navigation is essential to ensure smooth and adaptive experiences for IwCI and their support networks. Shared planning can play a key role, allowing caregivers and support workers to collaboratively plan outings and navigation tasks in real time \cite{Bae2024WayfindingIP}. To operationalize this, navigation systems should integrate shared digital artifacts—such as synchronized maps, editable plans, and shared calendars—that support real-time updates and bidirectional feedback. In addition, embedding customizable user profiles within these shared systems allows caregivers to store user preferences, frequently visited locations, and context-specific support needs, ensuring that planning remains consistent, accurate, and responsive to individual conditions. Empirical findings from both our survey and interview studies underscore this need, with participants emphasising that collaborative features can significantly reduce communication gaps and prevent planning errors (Section \ref{sec:preplan}). This recommendation reconceptualises collaboration as a coordinated multi-stakeholder requirement, bridging the traditionally individualistic design of navigation systems with socio-technical realities of shared care

\color{darkgreen}
\subsection{Implications for Software Engineering}

This study extends SE scholarship—particularly RE—by foregrounding the social and contextual dimensions of inclusivity in the design of navigation and assistive systems. The findings expose conceptual and methodological gaps in existing RE approaches and outline the directions for evolving SE practices to address the realities of cognitively diverse user populations.

\textbf{\textit{Rethinking Inclusive Requirements Engineering through Proxy Stakeholders}} Traditional elicitation techniques such as interviews, workshops, and prototyping \cite{coughlan2002effective, rizk2016requirements} often presume that users can easily articulate goals and preferences. However, for individuals with cognitive impairments (IwCI), cognitive and communication barriers can hinder such direct engagement, making conventional elicitation insufficient or inappropriate. Our findings call for a reconfiguration of requirements elicitation as a distributed and co-constructive process—one that actively involves proxy stakeholders such as caregivers, support workers, and allied health professionals. These proxies provide first-hand, sustained knowledge of user’ daily challenges and adaptive behaviors, offering essential insights that cannot be captured through direct elicitation alone. This reconceptualisation of elicitation as a networked process aligns with emerging inclusive design approaches that go beyond individual users to recognize broader ecosystems of care and support \cite{heumader2018requirements, ollerton2012ipar, lowdermilk2013user}. Proxy stakeholders do not merely supplement user voices—they are essential co-informants whose perspectives reflect both the lived interdependencies of support networks and the practical constraints of real-world technology use. Adopting this approach also implies a deeper commitment to the social model of disability, which reframes disability not as a user deficit but as the result of social and environmental barriers \cite{barnes2019understanding}. Within this framework, requirements are not merely functional specifications, but social commitments to accessibility and equity. This reorientation situates disability-inclusive software as a socio-technical construct, aligning with recent arguments for human-centered SE that integrates empathy and context awareness into technical decision-making \cite{grundy2020human}.

Our findings highlight the importance of proxy stakeholders in system use, yet identifying them systematically is a persistent challenge in software engineering. Despite their recognized value, many RE projects still focus primarily on primary users and developers \cite{sharp1999stakeholder, muller2021stakeholders}. This paper takes an initial step toward systematically engaging proxy stakeholders in the context of cognitive impairments. Drawing from our empirical findings, we propose a three-stage approach that includes the identification of appropriate proxies, strategies for meaningful engagement throughout the research process, and analytical techniques for interpreting proxy-derived data.

\textbf{\textit{Stage 1: Identifying Proxy Stakeholders.}} Begin by mapping out the care ecosystem surrounding the primary user (IwCI). Identify both informal proxies (e.g., family members, friends) and formal proxies (e.g., professional caregivers, therapists, support staff) who are involved in the user’s daily life. Previous research emphasizes that proxies can occupy various roles, ranging from providing direct information or context to confirming research results, making it essential to delineate who does what in the user’s life \cite{foley2019printer, lazar2017critical,lazar2018design}. In addition, clearly identifying the relationship of each proxy with the user (e.g., spouse, adult child, professional aide) is important for consistent terminology and helps anticipate their perspective. Second, Who qualifies as a proxy stakeholder can vary depending on the research or clinical context, ranging from those with shared lived experience to caregivers who are closely and consistently involved in the individual’s daily life \cite{bingham2017montreal,logsdon2002assessing}. For example, if researching navigation, include the caregiver who regularly assists the person in public places. Third, we should include a range of proxy types to capture diverse perspectives and avoid one-sided conclusions. Different proxies will have different expectations, levels of familiarity, and support styles with the user. For example, proxies, such as healthcare providers and caregivers, can have varying views on navigation preferences for individuals with IwCIs. Similarly, intermediaries such as teachers and parents of a child with autism can observe different elements of the child's experience \cite{alabdullatif2023use}. In practice, this means engaging both informal and formal supporters (e.g. a parent and a teacher; or a professional caregiver and a close relative).

\textit{\textbf{Stage 2: Engaging Proxy Stakeholders in the Research Process.}} When conducting interviews or workshops with proxies, frame your questions in concrete, familiar terms. For example, instead of asking a proxy abstractly about “\textit{navigation difficulties,}” you might pose a scenario like: \textit{“If you are with [User] in a busy shopping centre and they need to find the bathroom, how would you assist them?”}. Situated prompts improve data quality and align proxies with the user's perspective, revealing subtle insights about needs and coping strategies often missed by generic questions \cite{dai2021surfacing}. Second, engaging proxies requires careful ethical navigation to respect the primary user’s voice. One key step is to clearly distinguish when a proxy is speaking for the user versus speaking from their own perspective. Proxies, like support workers, may convey both their perception of the user's feelings and their professional opinion, which can unintentionally overshadow the participant's voice.

\textit{\textbf{Stage 3: Analysing Proxy-Derived Data.}} Once data collection is complete, a crucial analytic step is to annotate and organize the data by proxy characteristics. Tag each data segment (interview quotes, observations) with the type of proxy (formal vs. informal, family vs. professional) and their relationship to the user. By coding the data with proxy identity and role, you can later identify patterns or biases related to those roles. For example, formal caregivers (e.g., support workers) might often mention safety and protocol issues, while family members emphasize emotional aspects of the user’s experience. Analytical reliability improves when we stratify or compare data by proxy groups, rather than merging all proxy inputs as if it were identical. Secondly, use the unique insights gained from the proxies to uncover latent requirements that the primary users themselves might not explicitly articulate. Proxies, because of their supportive role, often observe unspoken anxieties, safety concerns, or workarounds that users develop, which are gold mines for design requirements that enhance system resilience and support. For instance, an individual with cognitive impairment might not mention their fear of getting lost, but a caregiver could reveal that they always insist the person carries a tracking device or avoids certain routes due to that concern. These proxy observations can point to features the user wouldn’t request but would benefit from. In a co-design study for children with communication challenges, researchers found that children struggled to express their issues or imagine solutions, making proxies (parents, teachers) crucial for identifying pain points and suggesting strategies \cite{alabdullatif2023use}.

\textbf{\textit{Operationalizing Inclusivity as a Software Quality Attribute}}
Inclusivity is increasingly being recognized not just as a design ethos but as a critical quality attribute of software. If the software we create excludes diverse populations – essentially marginalizing people who “don’t fit” the profile of its designers – it raises serious ethical issues and also business risks \cite{guizani2020gender}. Notably, the latest international standard ISO/IEC 25010:2023 has formalized this idea by adding Inclusivity as a software quality \cite{ISOIEC}. Emerging methods help software teams integrate inclusivity, like GenderMag, which systematically addresses gender-related inclusivity issues \cite{burnett2016gendermag}. Research indicates that using structured processes to address inclusivity issues can lead to significant improvements. For example, a “inclusivity debugging” method helped developers identify and resolve design flaws that affect diverse users, reducing inclusive bugs by 90\% for new users in an open-source project \cite{guizani2022debug}.

This demonstrates the value of treating inclusivity as an explicit and systematic concern within SE, which is especially critical in the context of navigation systems. These applications often place high cognitive and sensory demands on users—requiring them to interpret spatial information, follow instructions, and recover from errors in real-time \cite{golledge2003human,kitchin2002cognition}. However, the capacity to carry out these tasks varies widely between users, particularly those with cognitive or intellectual disabilities, age-related impairments, or neurodiverse processing styles \cite{prestopnik2000relations, schnitzler2015user, jamshidi2020wayfinding, kato2003individual}. Given this diversity, navigation tools must account for a wider range of user capabilities, making inclusivity not just a design preference, but a fundamental quality attribute. Across our recommendations, inclusivity emerges as a cross-cutting concern that should be operationalized and evaluated throughout the software development lifecycle. From a SE perspective, inclusivity should be part of quality models as a measurable feature. Developers can evaluate if navigation systems offer flexible options like simplified modes or alternative cues for diverse user needs. This \textit{"optionising for inclusivity" }enables tailoring without identity-based designs, focusing on capability diversity to enhance access and minimize exclusion.

%------------------Threats to Validity----------------------%
\section{Threats to Validity}
\color{darkblue}
\textbf{External validity.} 
 We gather data from healthcare professionals, support staff, and caregivers through semi-structured interviews, concluding our data collection with the 16th interview. The final interviews primarily served to validate our findings rather than to provide new information. In interview research, we employed purposive sampling technique, which emphasises extracting insights from \textit{representative samples instead of aiming for broad generalisability} \cite{hargittai2019internet, hoda2021socio}. Consequently, the qualitative findings of our study cannot be generalized to larger populations. Although the survey captured responses from a diverse international cohort, the concentration of participants in the United Kingdom and the dominance of health professionals may limit the generalizability of the findings to other geographic regions and professional roles. Future research should explicitly target a wider geographical representation, particularly from regions such as Asia, Africa, and South America, where different cultural contexts and infrastructure may yield additional information. A systematic comparison between regions would help validate whether our findings represent universal needs or are influenced by specific regional contexts. Certain cognitive impairments, such as depression and anxiety, were reported more frequently, while rarer conditions such as dysgraphia, Tourette’s syndrome, and schizophrenia were under-represented. This uneven distribution may limit our understanding of the needs of individuals with less common impairments. Although interview and survey studies can complement each other, a limitation arises due to the difference in cohorts: the survey involved participants from a broad international background, whereas the interviewees were primarily from Victoria, Australia. Such geographic and demographic differences might restrict the synthesis of insights. Additionally, the range of cognitive impairments discussed in the studies varied; these differences in conditions and experiences could impede drawing uniform conclusions across the datasets.\color{darkgreen} Another concern relates to the mixing of indoor and outdoor navigation contexts in our data collection. While our survey and interview instruments intentionally addressed both types of wayfinding environments, not all participants may have clearly differentiated their responses across these contexts. As such, some insights may reflect general impressions of navigational independence rather than environment-specific challenges. This ambiguity may limit the precision of certain design recommendations when applied to highly specialized indoor or outdoor settings. 
 
Reliance on caregiver proxies rather than direct IwCI participation limits the generalizability of the findings to end-user experiences. Although caregivers provide valuable information on longitudinal patterns, their perspectives may not fully capture: (1) immediate frustrations with interaction or (2) subjective preferences of IwCI. This aligns with known proxy-reporting biases in health related research \cite{todorov2000bias}. In addition, diversity in professional roles and support contexts may also affect the generalizability of the findings. Although we attempted to mitigate this through structured prompts and contextual phrasing in the survey, we acknowledge this variability as a threat to broader generalizability. Although our study aimed to identify cross-cutting challenges and inclusive design needs, the wide variation in cognitive conditions (e.g., autism, ADHD, intellectual disability) means that some findings may not generalize equally across all subgroups. As our focus was on synthesizing the shared patterns reported by carers, condition-specific needs may not be fully represented. This diversity highlights the importance of treating our findings as a foundation for inclusive design rather than as prescriptive solutions for all cognitive profiles. Future work is needed to explore condition-specific adaptations and evaluate how different user groups interact with configurable features. Furthermore, our study cannot identify the root causes of the wayfinding behaviors and competencies exhibited by IwCI \cite{gallagher2008predictors}. Although the conduct of caregivers and support workers influences the wayfinding activities of IwCI, these behaviors can also directly stem from the abilities or challenges IwCI faces. We only assessed general wayfinding activities both in our interview and in our survey study; however, wayfinding is a complex skill that relies on spatial, verbal, memorial, executive function, and social abilities \cite{golledge2003human,kitchin2002cognition}. Participants interact with various IwCI groups, who differ in dependence levels, ages, and types of cognitive impairment, and their roles with IwCI also vary. 
\color{darkgreen}

\textbf{Internal validity.} While our two-stage screening process implemented rigorous verification measures (including experience duration thresholds and response consistency checks), we acknowledge potential limitations inherent to online recruitment platforms. Despite Prolific's identity verification systems and our additional screening measures, some participants might misrepresent their caregiving experience. This could marginally affect the reliability of self-reported data. We also acknowledge that our interview questions were not formally validated through expert review or a dedicated internal or external pilot study. Instead, the first iteration of our interview process functioned as a formative phase in which two co-authors jointly facilitated interviews, conducted debriefs, and made minor refinements to the interview guide based on observed participant responses and emerging themes. Although this approach supported practical validation through real-time reflection and protocol adjustment, we recognize that it does not substitute for a structured validation process. As noted by \citet{creswell2016qualitative}, differences in the way participants of various cultural and professional backgrounds interpret qualitative prompts can affect the consistency of the data. Although we attempted to mitigate this by including a demographically diverse group of caregivers in the early phase, we acknowledge that potential biases in the framing and interpretation of the question remain a limitation. Future work should incorporate a formal validation process—either through expert review or a standalone pilot study—to strengthen the reliability and interpretive robustness of the interview protocol.
\color{darkblue}

We used the STGT technique for qualitative data analysis and promoted extensive team discussions to assess and improve our analyzes, findings, and presentation, thus minimizing potential biases. There might be chances of misinterpretations or misunderstandings between our questions and the responses of the participants. To address this and gain a clearer understanding of their views, we posed follow-up questions during the interviews and sought additional clarifications on their remarks. Although compensating interview study participants might raise concerns that participants provide misleading information to qualify for the study \cite{head2009ethics}, all participants recognize the importance of our research and are interested in learning about the study results, anticipating us to update them once preliminary results are available. Our survey and interview study relied on participants recounting past events. It would be beneficial to collect their reports of activities in real time instead of relying solely on memories. 

\color{darkblue}
\section{Related Work}

Few studies have specifically outlined the wayfinding requirements of IwCI \cite{gupta2020towards}. In contrast, a significant amount of research focuses on navigation for people with visual \textbf{impairment} \cite{somyat2018navtu, ganz2014percept, prerana2019stavi, paladugu2013supporting}, and numerous systematic literature reviews assess the current knowledge on designing navigation applications for users with visual impairments \cite{zahabi2023design, kuriakose2022tools, budrionis2022smartphone, khan2021analysis}. When we focus on IwCI, the numbers decrease significantly. The intricacies of the brain and individual differences make this field of research both difficult and intriguing \cite{braddock2004emerging}. 

Despite these challenges, literature reviews continue to highlight interesting projects and studies. Works such as \citet{beeharee2006natural}, \citet{fickas2008route}, and \citet{liu2009customizing} have investigated various methods to provide navigation instructions to IwCI, including the use of arrows, audio prompts, maps, and landmarks. Several research efforts explore a variety of guidance approaches and interface types for IwCI \cite{liu2006indoor, gomez2015design, chang2010autonomous}. Certain studies customize routes according to the' cognitive abilities of users, such as incorporating familiar landmarks to determine the optimal route \cite{hervas2013assistive}, and introducing an innovative route calculation system based on social cooperation \cite{holone2007users}. Furthermore, \citet{garcia2012should} discussed the challenge of differentiating between left and right using textual directions and the confusion caused by discrepancies between the actual scene and the image depicted, while \citet{fickas2010travelers} examined error recovery and discovered that IwCI encountered significant difficulties in articulating their location for reorientation.

Although most studies contribute significantly to understanding the performance of wayfinding in IwCI, laboratory studies alone cannot offer a comprehensive view of the daily challenges of wayfinding encountered by IwCI \cite{yang2018parent,bailey2010using, budrionis2022smartphone}. Furthermore, these studies do not measure many\textbf{ social and situational factors} that influence navigation activities \cite{yang2018parent}. Social factors, such as being around or engaging with others, can impact wayfinding, which is essentially a decision-making process where the actions or presence of individuals, present or past, visibly influence the decision-making involved \cite{dalton2019wayfinding,zacharias2001path,passini1981wayfinding}. Although other people are not always physically present while navigating or have not provided navigation guidance, it is evident that people, social groups, institutions, and cultural norms invariably impact the wayfinding process \cite{dalton2019wayfinding}. IwCI generally rely on the guidance of \textbf{caregivers, support workers, or health professionals}, who influence or assist in their decision-making process when navigating. Although much of your daily search involves multiple stakeholders, minimal research has explicitly focused on the social dynamics of tasks such as collaborative route planning and navigation to this point \cite{Bae2024WayfindingIP,yang2018parent,he2015collaborative}. Research on wayfinding has rarely explored it as a societal issue \cite{dalton2019wayfinding}, often focusing instead on the challenges faced by individuals, partly due to the additional complexity of evaluating decision-making processes involving multiple individuals and their possible emergent interactions \cite{Bae2024WayfindingIP, yassin2021others}. An additional consideration is the increasing application of virtual reality technologies as experimental instruments and settings in wayfinding research, which generally do not include the presence of other "people" \cite{meilinger2008working,mengue2011route,chaudary2023teleguidance}.

%------------------Conclusion----------------------%
\section{Conclusion}
This study advances human-centric software engineering by formally recognizing the critical role of proxy stakeholders—carers, support workers, and health professionals who actively shape how individuals with cognitive impairments (IwCI) navigate real-world environments. Using a mixed-method approach grounded in qualitative insights, we identified design requirements that extend beyond the individual user to encompass the relational and contextual dynamics of supported navigation. Our findings reveal a critical limitation in current requirements engineering (RE) practices: by prioritizing direct user input, they often overlook socially embedded support systems essential to the success of many users’. Our primary methodological contribution is the formalization of the proxy stakeholder—an actor distinct from an indirect stakeholder, who not only is affected by a system, but actively mediates, interprets, and scaffolds its use. We provide a practical framework for identifying, engaging, and translating the input of these proxies, thereby advancing more inclusive and effective RE. Together, these methodological and empirical contributions support more ecologically valid and ethically grounded RE practices. Our design recommendations specifically promote the development of customizable, collaborative, proxy-aware features that align with the social model of disability and respond to the real-world contexts of IwCI.

By centering proxy stakeholder perspectives, this approach challenges traditional user-centered design paradigms and offers a pathway toward more socially aware requirements engineering. Future research should focus on three key areas: (1) operationalizing these design recommendations within specific technology platforms such as mobile navigation apps and assistive wearables, (2) conducting longitudinal evaluations of proxy-informed systems in real-world deployments, and (3) extending the proxy stakeholder framework to other domains where vulnerable populations rely on support networks. Ultimately, by bringing proxy stakeholders from the margins to the center of the design process, this research lays the foundation for a new class of software systems—those that are not just inclusive by design, but empathetic and adaptable by nature, truly reflecting and supporting the lived experiences of cognitively diverse populations.

\color{black}

\begin{acks}
This work is partially supported by the Australian Research Council Laureate Fellowship FL190100035 and the Monash Assistive Technology (MATS) SEED grant. We would like to acknowledge Ilianna Ginnis, Kirsten Ellis, Kylie Edgar, and Janet Lloyd-McNelis for helping us recruit participants for the user study. Special thanks to all the participants for sharing their valuable perspectives and experiences. 
\end{acks}

%%
%% The next two lines define the bibliography style to be used, and
%% the bibliography file.
\bibliographystyle{ACM-Reference-Format}
\bibliography{reference}

%%
%% If your work has an appendix, this is the place to put it.
\appendix
\section{Survey: Stage one} \label{app:stageone}
\begin{footnotesize}
\begin{enumerate}
    \item In what age group are you?
    \item To which gender identity do you most identify as?
    \item What's your role? Choose from:Health Professionals, Caregivers, Support workers
    \item What's your health profession? (if Health Professionals is selected for the previous question)  Medical practitioner, Nurse, Occupational therapist, Psychologist, Speech pathologist, Physiotherapist, Other (Please specify)
    \item Have you worked with individuals with cognitive impairment (IwCI) as a part of this role? If Yes, how many years?
    \item Enter your PROLIFIC ID (Please note that this response should auto-fill with the correct ID)
\end{enumerate}

\end{footnotesize}

\section{Survey: Stage two} \label{app:stagetwo}
\begin{footnotesize}
\begin{enumerate}

    \item In which country do you currently reside?
    \item Could you please specify the particular cognitive impairment or diagnoses of the individuals you have cared for? (Select all options apply) 
    \begin{itemize}
        \item Autism 
        \item ADHD 
        \item Anxiety 
        \item Depression 
        \item Obsessive Compulsive Disorder (OCD) 
        \item Bipolar 
        \item Borderline Personality Disorder (BPD) 
        \item Post Traumatic Stress Disorder (PTSD) 
        \item Dyslexia
        \item Epilepsy
        \item Tourette's 
        \item Down Syndrome
        \item Cerebral Palsy
        \item Dysgraphia
        \item Dyscalculia
    \end{itemize}
    \item What is your Prolific ID? Please note that this response should auto-fill with the correct ID
    \item Challenges in Wayfinding for cognitive impairment 
    Kindly share your insights based on your experience in working with/caring for IwCI.
    \begin{itemize}
        \item What are the common public spaces IwCI visit?
        \item How would you rank their ability to navigate these spaces independently? (1- unlikely, 5 very likely)
    \end{itemize}
    \item Challenges in Wayfinding when travelling with Cognitive Impaired. Kindly share your insights based on your experience in visiting Public Spaces with IwCI.
    \begin{itemize}
        \item What sensory difficulties make it challenging for IwCI to visit these spaces? Select all options apply
        \begin{itemize}
            \item Auditory (e.g., Shopping centres with crowded areas, music.) 
            \item Smell (e.g., Strong odours from perfumes or exposure to unpleasant smells in hospitals or public spaces.)
            \item Temperature (e.g., Inconsistent heating or cooling.)
            \item Visuals (e.g., Overly bright lighting or sudden changes in lighting conditions.)
            \item Motion (e.g., Difficulty with spaces that induce motion, such as revolving doors or moving walkways.)
        \end{itemize}
        \item What other challenges do they face in visiting these spaces?
        \begin{itemize}
            \item Crowds challenges
            \item Memory challenges ( e.g., remembering directions or locations.)
            \item Communication challenges (e.g., communication and social cues, potentially leading to misunderstandings.)
            \item Time challenges (e.g., challenges in estimating time and adhering to schedules.)
            \item Difficulty with decision-making
            \item Map-reading challenges
            \item Other (Please specify)
        \end{itemize}
        \item In terms of facilities, what information do you and the IwCI for need to know in visiting these spaces?\begin{itemize}
            \item Location of accessible toilets
            \item Location of lifts
            \item Location of accessible parking
            \item Location of accessible seating Areas
            \item Location of staircase locations
            \item Nearby cafes or food courts
            \item Nearby transportation hubs
            \item Quiet areas (if any)
            \item Other (Please specify)
        \end{itemize}
        \item How likely are you to pre-plan your trip with a IwCI to a complex indoor space? (1- unlikely, 5-very likely)
        \item Can you list the steps you follow in pre-planning? (e.g. refer to the building plan/map, share the plan with the people with cognitive impairments, warn them the day before, etc.) If not applicable, leave NA
        \item If you couldn’t pre-plan, how do you find your way in a very complex space with the IwCI? (Select all options apply)
        \begin{itemize}
            \item Ask information desk
            \item Ask Fellow Visitors
            \item Follow Signage
            \item Refer to the wayfinding map on the kiosk
            \item Refer to the map on the phone
            \item Other (Please specify)
        \end{itemize}
        \item During these visits, in using digital maps (on kiosks /the web/ the phone), what features have you liked most? If not applicable, leave NA
        \item In using these maps, what frustrations have you faced? If not applicable, leave NA
        \item If you could improve/change these maps in any way you want, what changes would you make to the map so that it helps you more on these visits with IwCI? Select all options apply
        \begin{itemize}
            \item Changes to pre-planning
            \item Changes to Navigation instructions: 
            \item Changes to Navigation visuals:
            \item Changes to map colors: 
            \item Changes to Map icons: 
            \item Changes to text labels:
            \item Other changes: 
        \end{itemize}
        \item If the person with cognitive impairment are to use these maps, what features should a digital map have?

    \end{itemize}
    
\end{enumerate}
 
\end{footnotesize}
\end{document}